\documentclass[aip,jcp,floatfix,citeautoscript,reprint]{revtex4-1}
\setcitestyle{super}

\usepackage{graphicx}
\usepackage{setspace}[1.5]
\usepackage{amsmath, amsthm, amssymb}
\usepackage{xfrac}
\usepackage{dcolumn}
\newcommand{\etal}{\textit{et al.}}

\newcommand{\rsc}{r$_{\text{s}}$}
\newcommand{\s}{$\sim$}

\DeclareMathOperator\erf{erf}

\begin{document}

\date{\today}

\title{Exploring water adsorption on isoelectronically doped graphene using
alchemical derivatives}

\author{Yasmine S. Al-Hamdani}
\affiliation{Thomas Young Centre and London Centre for Nanotechnology,
  17--19 Gordon Street, London, WC1H 0AH,
  U.K.}
\affiliation{Department of Chemistry, University College
  London, 20 Gordon Street, London, WC1H 0AJ, U.K.}

\author{Angelos
  Michaelides} \affiliation{Thomas Young Centre and London Centre for
  Nanotechnology, 17--19 Gordon Street, London, WC1H 0AH,
  U.K.}  \affiliation{Department of Physics and Astronomy, University
  College London, Gower Street, London WC1E 6BT,
  U.K.}

% \author{Dario
%   Alf\`{e}} \affiliation{Thomas Young Centre and London Centre for
%   Nanotechnology, 17--19 Gordon Street, London, WC1H 0AH,
%   U.K.}  \affiliation{Department of Earth Sciences, University College
%   London, Gower Street, London WC1E 6BT, U.K.}

\author{O. Anatole von Lilienfeld} 
  \email{anatole.vonlilienfeld@unibas.ch}
  \affiliation{Institute of Physical
  Chemistry and National Center for Computational Design and Discovery
  of Novel Materials (MARVEL), Department of Chemistry, University of
  Basel, Klingelbergstrasse 80 CH-4056 Basel,
  Switzerland}

\begin{abstract} The design and production of novel 2-dimensional materials has
seen great progress in the last decade, prompting further exploration of the
chemistry of such materials. Doping and hydrogenating graphene is an
experimentally realised method of changing its surface chemistry, but there is
still a great deal to be understood on how doping impacts on the adsorption of
molecules. Developing this understanding is key to unlocking the potential
applications of these materials. High throughput screening methods can provide
particularly effective ways to explore vast chemical compositions of materials.
Here, alchemical derivatives are used as a method to screen the dissociative
adsorption energy of water molecules on various BN doped topologies of
hydrogenated graphene. The predictions from alchemical derivatives are assessed
by comparison to density functional theory. This screening method is found to
predict dissociative adsorption energies that span a range of more than 2 eV,
with a mean absolute error $<0.1$ eV. In addition, we show that the quality of
such predictions can be readily assessed by examination of the Kohn-Sham highest
occupied molecular orbital in the initial states. In this way, the root mean
square error in the dissociative adsorption energies of water is reduced by
almost an order of magnitude (down to $\sim0.02$ eV) after filtering out poor
predictions. The findings point the way towards a reliable use of first order
alchemical derivatives for efficient screening procedures. \end{abstract}
\maketitle

\section{Introduction} Recognising the enormous number of ways in which elements
can be combined is both exciting and daunting in the search for more efficient,
more sustainable, and safer materials for medical, engineering, and catalytic
applications. High throughput screening in computational chemistry, otherwise
known as virtual screening, is paving the way for materials discovery across
academic and industrial research. There are various ways to screen through
materials (see e.g.
Refs.~\citenum{curtarolo2013,pyzer2015,pfeif2016,faber2016,weymouth2014,
cerqueira2015}). One particularly noteworthy example in catalysis was the study
of Greeley \etal\ which involved the computational screening of 700 binary
surface alloys to find a material with high activity for H$_2$ evolution
\cite{greeley2006}. The computational screening lead to the discovery and
subsequent synthesis of BiPt which showed comparable activity to pure Pt
experimentally.

We focus on an area of widespread interest that is, dissociative molecular
adsorption on 2-dimensional substrates. In particular, graphene and hexagonal
boron nitride (h-BN) are nearly isostructural materials with emerging
applications in industry, including in
catalysis\cite{deng2016catalysis,li_2013semi,sun_2011,li2011highly,zheng2013two,kong2014doped,
machado2012graphene,fan2015multiple}. However, an important challenge in using
graphene for catalysis, is overcoming its inertness. There are a number of ways
to facilitate reactions at the surface of graphene such as using metal
substrates\cite{machado2012graphene,fan2015multiple,yao2014graphene,Grosjean2016,He2017,Lattelais2015,
Altenburg2015} to electronically dope graphene and in-plane doping of graphene
with other elements\cite{al2016tuning,zheng2013two,kong2014doped,Sheng2011}. For
instance, pristine graphene has been shown to be inert to the dissociative
adsorption of water whereas, BN doped and hydrogenated graphene is far more
likely to dissociate water\cite{al2016tuning}. Hydrogenating graphene breaks the
large delocalized $\pi$ network of electrons in graphene, which is key to its
inertness\cite{liao2014jacs,kong2014jpcm,liu2015nsr}. The hydrogenation of
graphene has been extensively studied in experiments, with a number of methods
of production (see \textit{e.g.} Refs.
\citenum{Elias2009,jaiswal2011,luo2009,wojtaszek2011}). In addition, doping
graphene isoelectronically with BN atoms further facilitates the adsorption of
molecules by forming stronger covalent bonds with
adsorbates\cite{al2016tuning,Grosjean2016}. The in-plane BN doping of graphene
has also been realised experimentally in recent
years\cite{Gong2014,Ci_2010,Liu_2013,synBNDG}, with increasing control over the
doping process such that, nanometre-scale domains can be
produced\cite{Liu_2013,Gong2014}, as well as separated B and N atoms in the
graphene surface\cite{synBNDG}.
% the point
Facilitating adsorption processes in such ways is vital for these materials
to become energy efficient and applicable on a large scale.  Here, we
investigate how isoelectronically doping with BN away from the adsorption site,
affects the dissociative adsorption energy of water on graphene.

Considering that computational molecular adsorption studies on graphene
typically involve unit cells containing 30-50 carbon atoms, there are hundreds
of ways to arrange a pair of boron and nitrogen atoms in such a unit cell (after
accounting for redundancies by symmetry). However, the isoelectronic nature of
doping in this study, and the proximity of boron, nitrogen, and carbon in the
periodic table, can be utilized for efficient approximate screening schemes.
Specifically, we can look to alchemical derivatives in density functional theory
(DFT)\cite{von2006molecular,Yang1985,Parr1984,geerlings2003conceptual}. This
method relies on exploiting the information encoded in the averaged
electrostatic potential at each atom, which is analogous to the first order
alchemical derivative, readily available after any self-consistent field (SCF)
calculation. This and similar conceptual DFT has been discussed comprehensively
in some
contributions\cite{von2006molecular,von2013first,geerlings2014conceptual}, and
later in Section \ref{background} we give a brief introduction to the method
employed.
% examples
Note that alchemical derivatives have been used previously to predict various
properties such as, intermolecular energies\cite{von2007jctc}, HOMO
eigenvalues\cite{von2010jcp}, reaction energies\cite{sheppard2010alchemical},
doping in benzene\cite{homoBenz,balawender2013exploring}, covalent
bonds\cite{chang2016fast}, and binding in alkali halide crystals,
\cite{solovyeva2016rapid} or transition metals
\cite{moritz2016jcp,Weigend2014,weigend2004atom}.

In this study, the first order alchemical derivative is used to predict the
dissociative adsorption energy of water on BN doped graphene, with doping
occuring at different sites in the substrate. The predicted energies are
compared to explicitly calculated energies to reveal the quality of predictions
and to identify any outliers. Further, it is shown that outliers can be
identified without additional calculations by simply using $\rho_{HOMO}$ of the
initial state. The study begins with a description of the methods and the system
setup in Section \ref{background}, followed by the results of alchemical
predictions in Section \ref{alchemRESULTS}. After identifying the main trends,
further questions about the procedure and implications for water adsorption are
discussed in Section \ref{discussion} before concluding in Section
\ref{alchemconc}.

\section{Methods}\label{background} Let us begin with a brief background,
followed later by details of the system setup and calculations. Firstly, any
point in chemical compound space can be referred to as a discrete chemical
thermodynamic micro-state.  Within DFT, such a state is defined by the charge
density, which results from solving an equivalent of Schr\"odinger's equation
for a given proton distribution $Z(\mathbf{r})$ and number of electrons $N_e$.
As such, $Z(\mathbf{r})$ and $N_e$ can also be seen as extensive particle
variables in a molecular grand-canonical ensemble\cite{von2006molecular}. The
mutation of a chemical thermodynamic system into another can be achieved by
thermodynamic integration with respect to a switching parameter $\lambda$. The
parameter $\lambda$ simply tracks the change from the initial state to the final
state. A converged integration would require sampling intermediate $\lambda$ and
hence, several DFT calculations. Instead here, this mutation is approximated,
using a Taylor expansion around the initial system and $\lambda$,
\begin{equation}\label{taylor} E(\lambda=1)=E^0 +
\partial_{\lambda}E^{0}\Delta\lambda+\frac{1}{2}\partial^{2}_{\lambda}E^{0}\Delta\lambda^{2} +
\dotsc,   \end{equation}
where $\lambda=0$ corresponds to the initial system, $\lambda=1$ corresponds to
the target system and hence, $\Delta\lambda=1$. Indeed it is not given that the
first order term in Eq. \ref{taylor} is always predictive. However, it has been
observed that for relative energies, such as the adsorption energy for instance,
higher order terms can cancel out resulting in useful predictions of
properties\cite{von2007jctc,von2010jcp,sheppard2010alchemical,homoBenz,balawender2013exploring,chang2016fast,
solovyeva2016rapid,moritz2016jcp,Weigend2014,weigend2004atom}. Importantly, as
we see below, the first order term in Eq. \ref{taylor} can be evaluated from a
single DFT calculation of the initial state. In general, the first order term
($\partial_{\lambda}E^0$) includes the variance of the energy with changes in
the proton density, the nuclear positions, and the number of electrons. However,
here we consider the isoelectronic doping of a graphene sheet with fixed atomic
positions, and later this is shown to be a good approximation in the system
considered here. Terms involving changes in atomic positions \{$\mathbf{R}_I$\}
and $N_e$ can therefore be neglected leaving us with the electronic
contribution,
\begin{eqnarray}
\label{fod} \partial_{\lambda}E & = & \sum_I \frac{\partial E}{\partial Z_I} \frac{\partial Z_I}{\partial \lambda} \\
& = & \sum_I \int d{\bf r} \frac{\rho({\bf r}) \erf[\sigma
|{\bf R}_I - {\bf r}|]}{|{\bf R}_I - {\bf r}|} 
\frac{\partial Z_I}{\partial \lambda } 
\;\; = \;\; \sum_I
\mu_I 
\frac{\partial Z_I}{\partial \lambda }  \nonumber
\end{eqnarray}
where the variation of the energy with respect to a small change in nuclear
charge ($Z_I$), damped by the error-function because of the lack of intranuclear
repulsion, is known as the alchemical potential
$\mu_I$\cite{von2005variational}.  This is referred to as the alchemical
potential, rather than the electrostatic potential, since it quantifies the
first-order energy change as a result of an ``alchemical'' infinitesimal
variation in proton number at an atomic site.  When deviating from the
transmutating atom's position, the alchemical potential becomes very similar to
the electrostatic potential, $\bar{V}_{ESP}(\mathbf{r})$. For practical reasons
we note that the average electrostatic potential at each atom (including the
nuclear contributions omitted in Eq. \ref{fod}) - or alchemical potential - is
readily available at the end of the SCF cycle in the widely used Vienna
\textit{Ab-Initio} Simulation Package (VASP)\cite{vasp1,vasp2,vasp3,vasp4}.
Hence, we can easily evaluate the first order alchemical perturbation based
approximation of the energy of {\em any} doped system from the information ({\em
i.e.} $\bar{V}_{ESP}(\mathbf{r})$) provided in a single DFT calculation
containing all of the atoms relevant to the doping process.

Not surprisingly, however, the quality of first order based predictions can vary
significantly, and it is expected that the second order derivative in Eq.
\ref{taylor} can improve the accuracy of predictions\cite{chang2016fast} by
introducing some response properties of the system. For example, the second
order term includes variation of the alchemical potential with respect to
nuclear charge,
\begin{equation} \partial_{Z_I} \mu_I = \int d{\bf r} \frac{\erf[\sigma |{\bf
R}_I - {\bf r}|]}{|{\bf R}_I - {\bf r}|} \partial_{Z_J} \rho({\bf r}),
\end{equation}
where $\partial_{Z_I} \rho({\bf r})$ corresponds to the electron density
response to varying the nuclear charge at the doping atom $I$. There are various
ways to calculate the electron density's response which involve further
computational effort, for this work we merely wish to estimate it in a
qualitative fashion. As such, we find it useful to assume the existence of a
correlation between the actual response and the Pearson's local softness of the
atom in the molecule, as measured by the local density of the highest occupied
molecular orbital (HOMO) for electrophiles (such as protons),
$\rho_{HOMO}$~\cite{pearson1986pnas}.

\subsection{Technical details and system setup}\label{alchemMETHOD} The
dissociative adsorption of a water monomer on boron nitride doped graphene
(BNDG) was calculated using DFT and VASP 5.3.2\cite{vasp1,vasp2,vasp3,vasp4}.
VASP uses plane-wave basis sets and projector augmented wave (PAW) potentials
\cite{PAW_94,PAW_99} to model the core region of atoms. The PBE
exchange-correlation functional\cite{PBE} is used throughout along with PBE PAW
potentials and a plane-wave energy cut-off of 500 eV. Earlier work has shown
that similar trends in terms of water dissociation are obtained with PBE, the
hybrid B3LYP\cite{b3lypA,b3lypB,b3lypC,b3lypD} functional, and the dispersion
inclusive optB86b-vdW\cite{vdwDF,B86,vdw_opt11} functional.\cite{al2016tuning}
The dissociative adsorption energy of water was found to be converged to 0.001
eV with a plane-wave energy cut-off of 500 eV when tested up to 800 eV. A
$(7\times7)$ unit cell of graphene is used, with four carbon atoms replaced by
two boron and two nitrogen atoms. The dissociative adsorption energy of water is
already converged with a ($5\times5$) unit cell but using a larger cell provides
more pathways for alchemical mutation of atoms. The separation between periodic
images of the substrate in the z-direction is 10 \AA; this achieves convergence
of the adsorption energy of water to within 0.004 eV compared to a z-direction
separation of 30 \AA. Reciprocal space was sampled with up to $7\times7\times1$
\textbf{k}-points and the adsorption energy was found to be converged within
0.05 eV at the $\Gamma$-point. Hence, all calculations reported here were
performed at the $\Gamma$-point.

\begin{figure}[ht] \centering
\includegraphics[width=0.5\textwidth]{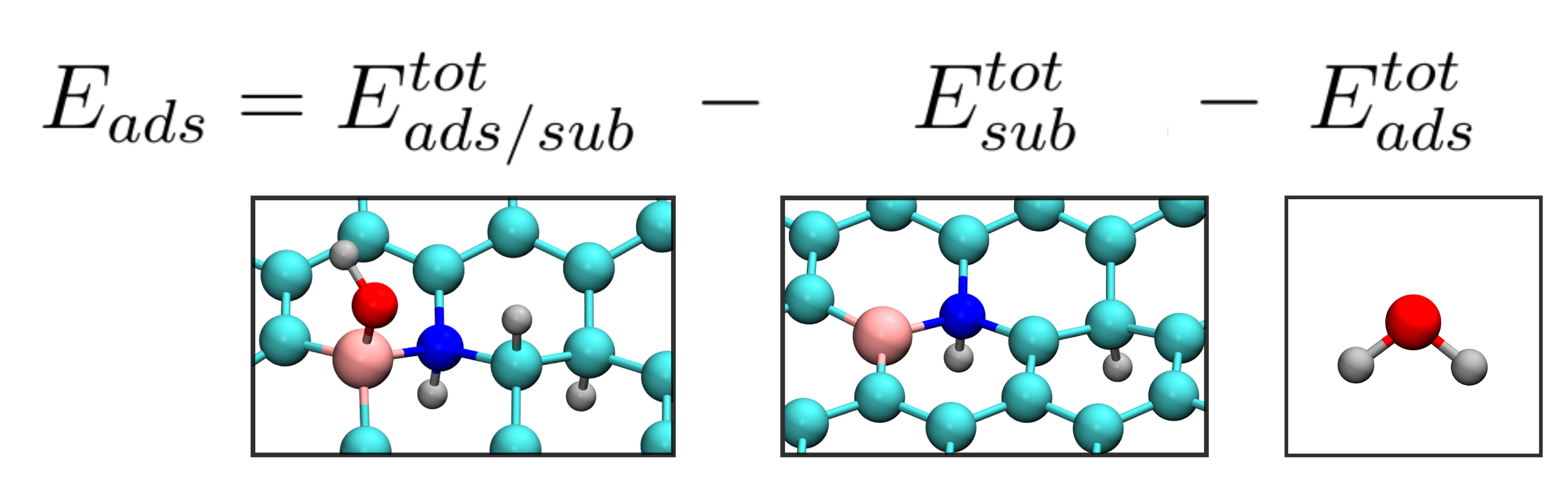} \caption{Adsorption energy
$E_{ads}$ defined as the difference between the adsorption system
($E_{ads/sub}^{tot}$), and the substrate with two hydrogen atoms adsorbed
($E_{sub}^{tot}$) and the gas phase water molecule ($E_{ads}^{tot}$). Water is
dissociatively adsorbed on the opposite side of the sheet to the hydrogen atoms.
Carbon is in light blue, nitrogen in dark blue, boron in pink, oxygen in red and
hydrogen in grey.}\label{figure1} \end{figure}

%% \subsection{System Setup}
The adsorption site in the substrate contains a pair of BN atoms in the surface
and two adsorbed hydrogen atoms, as shown in Fig.~\ref{figure1}. Doping and
hydrogenating in this way has been shown previously to make the surface more
reactive towards the dissociative adsorption of water\cite{al2016tuning}.
Importantly, atoms other than carbon at the adsorption site remain unchanged and
are not involved in any transmutations. The dissociative adsorption energy is
defined as, \begin{equation}
E_{ads}=E^{tot}_{ads/sub}-E^{tot}_{sub}-E^{tot}_{ads}\label{IE-1} \end{equation}
where $E^{tot}_{ads/sub}$ is the total energy of the adsorption system,
$E^{tot}_{sub}$ is the total energy of the substrate (with two hydrogen atoms
adsorbed), and $E^{tot}_{ads}$ is the energy of the intact water molecule in the
gas phase. 

\begin{figure}[ht] \centering
\includegraphics[width=0.50\textwidth]{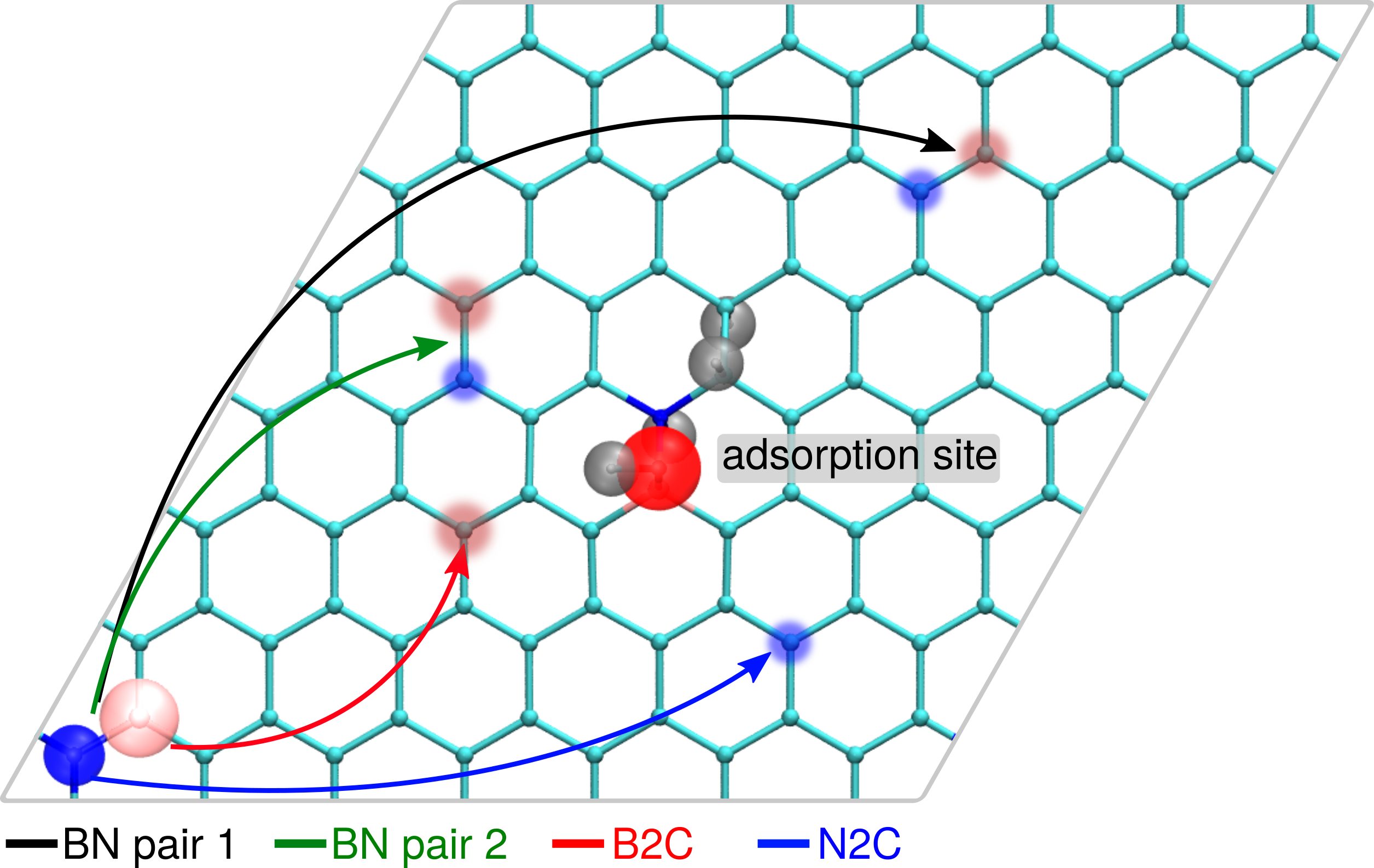}
\caption{$(7\times7)$ unit cell of BN doped graphene with water and hydrogen
atoms adsorbed. The substrate is doped with two pairs of BN atoms. The central
BN pair is not involved in transmutation and all atoms at this adsorption site
remain unchanged. The colored lines show example transmutation paths for BN pair
1 (black), BN pair 2 (green), B2C (red), and N2C (blue).}\label{figure2}
\end{figure}

Four types of alchemical mutation routes between carbon, boron, and nitrogen are
considered here, illustrated in Fig.~\ref{figure2}. There are a set of paths
associated with each route, where a path defines the starting and final states
for a given transmutation.  The initial state in each path contains a pair of BN
atoms near the edge of the unit cell which can be involved in transmutation.
Note that in all four alchemical routes, the graphene sheet is also hydrogenated
and contains a second pair of BN atoms at the dissociation site, but these
particular dopants are excluded from alchemical mutation. In two types of
routes, referred to as BN pair 1 and BN pair 2, a pair of BN atoms are
transmutated to different sites across the graphene sheet, as illustrated with
examples in Fig.~\ref{figure2}.
%% The difference between BN pair 1 and 2 is the position of the
%% unchanging BN atoms at the dissociation site with respect to the
%% transmutating BN pair; the two schemes are distinguishable due to
%% the two sublattices of graphene.
These two routes are distinguishable due to the existence of two sublattices
within graphene.  In BN pair 1, the transmutating BN atoms occupy the same
sublattice in graphene as the unchanging BN atoms at the dissociation site.
Whereas in BN pair 2, the transmutating BN atoms occupy the other sublattice.
The third type of route, B2C, refers to alchemical changes involving only the
boron atom.  Similarly N2C refers to the swapping of carbon atoms with nitrogen
while keeping the boron atom fixed.  In each type of route there are 94 possible
paths for this unit cell size, such that we have validated a total of 376 paths
for this study.  Note that only two single point DFT calculations are needed to
make alchemical predictions for a set of 94 paths.

Thanks to the geometrical similarity of graphene and hexagonal boron nitride,
doping graphene with BN atoms has a small impact on the structure. The largest
change in bond lengths upon relaxation of target systems was seen for
boron-carbon bonds, which changed by up to 0.06 \AA. The energy of relaxation
gained from this is up to $\sim0.3$ eV and does not alter the trends observed.
This makes fixing the geometry in all calculations a reasonable approximation to
begin with. 

\section{Results}\label{alchemRESULTS}
\begin{figure}[ht]
\centering \includegraphics[width=0.45\textwidth]{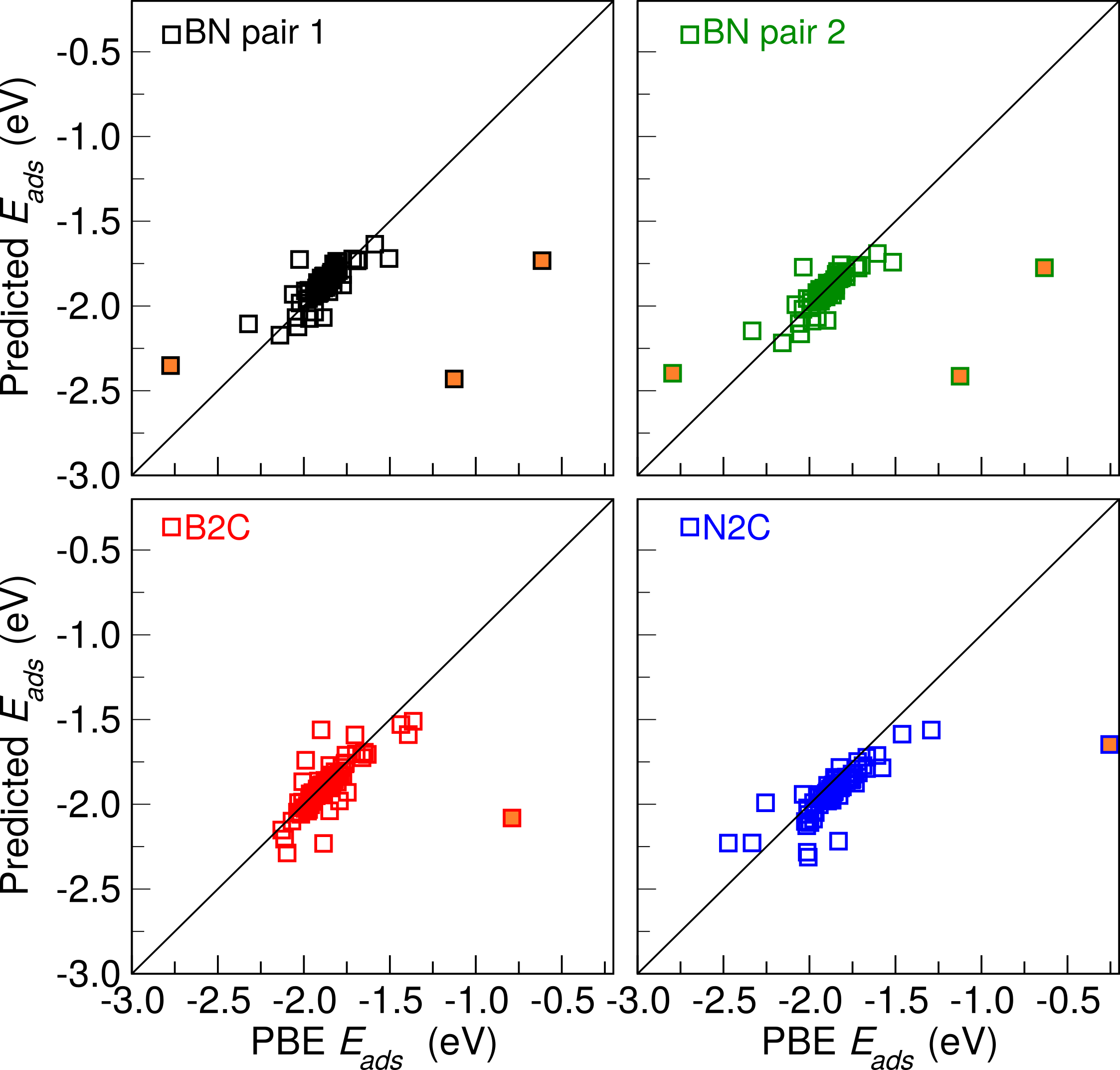}
\caption{Scatter plots of the PBE adsorption energies against alchemically
predicted adsorption energies for each path in eV. (a) BN pair 1 in black. (b)
BN pair 2 in green. (c) B2C in red. (d) N2C in blue. Clear outliers are
indicated by filled orange squares.}\label{figure3} \end{figure}

The PBE energy of water dissociation has been calculated for each transmutation
path without relaxing the positions of the atoms and compared with the alchemically
predicted dissociation energy. Fig. \ref{figure3} shows scatter plots comparing
these energies, for each alchemical route. It can be seen that the PBE
adsorption energies range from $-0.3$ to $-2.8$ eV, revealing that the precise
location of the dopants has a significant impact on the reactivity of the active
site.
%% chemistry
The large range of adsorption energies for seemingly similar surfaces can be
understood in terms of two chemical effects from the doping boron and nitrogen
atoms, namely, resonance and induction. In the former, the non-bonding valence
electrons of nitrogen partake in $\pi$ conjugation with p-states on carbon
atoms. This has a long-range impact on the electron density of the surface and
therefore the reactivity of the adsorption site. Second, the difference in
electronegativity between boron, carbon, and nitrogen atoms leads to local
inductive effects and this is likely to have a particularly large impact when
the doping atoms are near the active site.
Upon considering how well the alchemical derivatives capture this behaviour, it
can be seen that the majority of predictions is good. There are, however, a
number of outliers resulting in a poor R$^2$ correlation coefficient of 0.14 for
the BN pair 1 route. The R$^2$ coefficients for the other alchemical routes are
similarly unimpressive between 0.17-0.49, and in all cases there are clear
outliers. In addition, the few outliers correspond to configurations with either
the most or least favorable adsorption energies -- and the predictive power of
the first order alchemical derivatives is worse for the outliers with the less
favorable adsorption energies. These potentially interesting configurations are
considered in more detail in Section \ref{discussion}, but first it is important
to avoid predicting misleading trends for the outliers. It follows that for an
effective screening process, it would be better to identify outliers without
further computational cost. In the following section it is demonstrated how that
is possible using the HOMO in the initial states.

\subsection{Filtering outliers using highest occupied molecular orbitals}
\begin{figure}[ht] \centering
\includegraphics[width=0.50\textwidth]{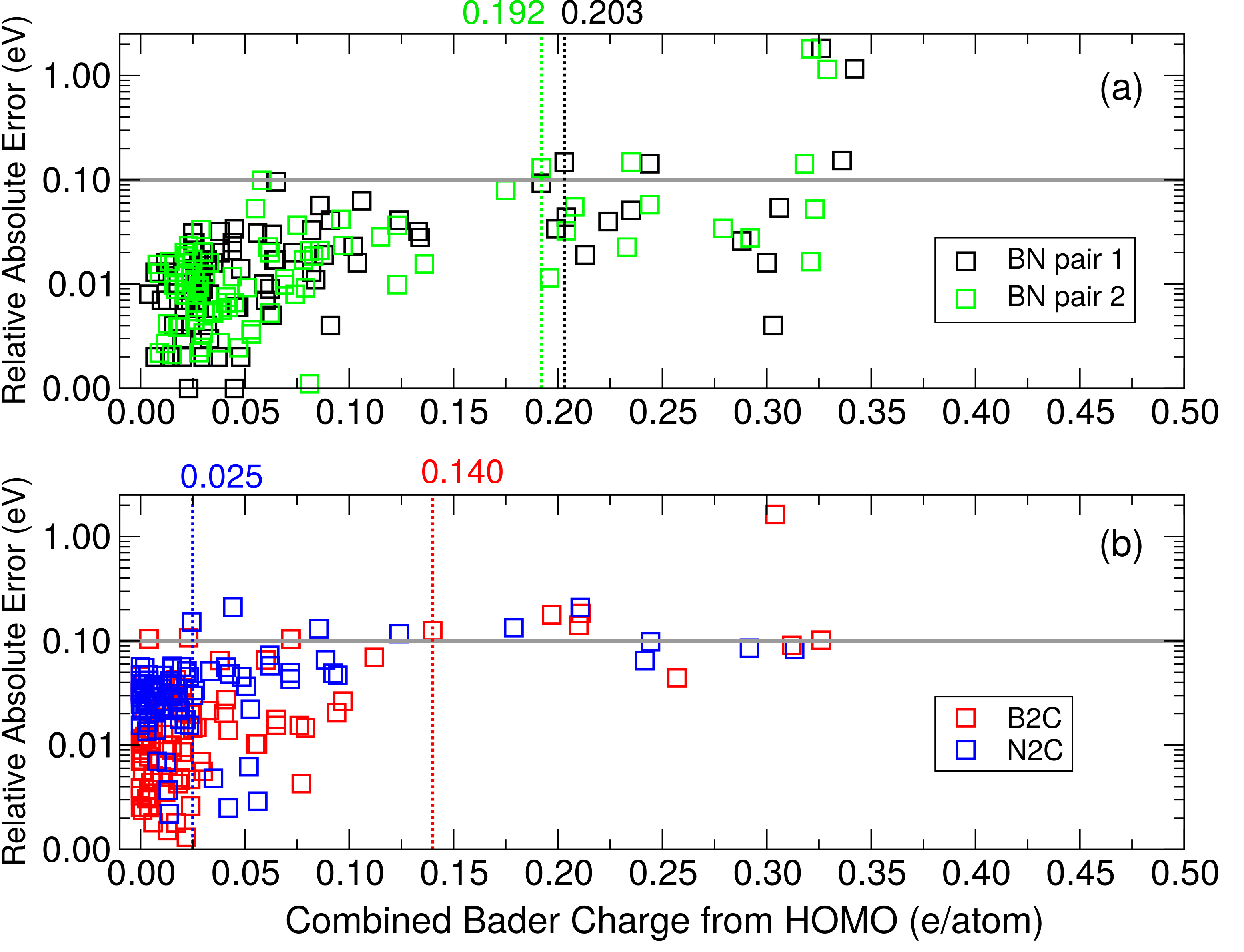} \caption{The relative
absolute error and combined HOMO charge is shown for each path. Top panel
includes BN pair 1 and 2, whilst lower panel includes B2C and N2C. Grey
horizontal lines indicate the threshold RAE value at 0.1. The vertical dashed
lines indicate the corresponding charge cut-offs for each route. These are used
to distinguish between paths which are referred to as mesomerically active
(higher than the charge cut-off) and passive (lower than the charge cut-off).
}\label{figure4} \end{figure}

Let us first consider doped graphene in which the substituent atoms and dopants
have a mesomeric effect on the electronic structure of the surface,
\textit{i.e.} they have either an electron withdrawing or electron releasing
impact. This effect resonates across the surface giving rise to mesomerically
active and passive sites. The mesomeric role of atoms can be probed using a
Bader charge density partition\cite{bader1990atoms} per atom of the HOMO charge
density, which indicates the prominence of the HOMO at a given atom site. Atoms
with charge density above a chosen cut-off value in the HOMO are considered
mesomerically active, and those under are mesomerically passive. For a given
path, the charge at the sites of mutation in the initial state can be summed to
obtain a measure of the extent of mesomeric activity. This combined Bader charge
and the corresponding relative absolute error (RAE) for each path is shown in
Fig.~\ref{figure4}. It can be seen that most paths have a RAE less than 0.01,
whilst those which have substantial errors also have large HOMO charges
associated with them. As a result, the partitioned HOMO charge can be used to
eliminate the outliers. Note that the correlation is not direct, there are some
paths with a high associated HOMO charge but small errors.
%% \begin{figure}[ht]
%% \centering
%% \includegraphics[width=0.45\textwidth]{./figure3.png}
%% \caption[Charge density plot of the Kohn-Sham highest occupied
%%   molecular orbital for the initial states in screening.]{Charge
%%   density plot of the Kohn-Sham highest occupied molecular orbital
%%   (HOMO) for the initial states in BN pair 1 scheme. An isovalue of
%%   0.005 eV/\AA$^3$ was used for the figures.}\label{figure3}
%% \end{figure}

\begin{figure}[ht] \centering
\includegraphics[scale=0.4]{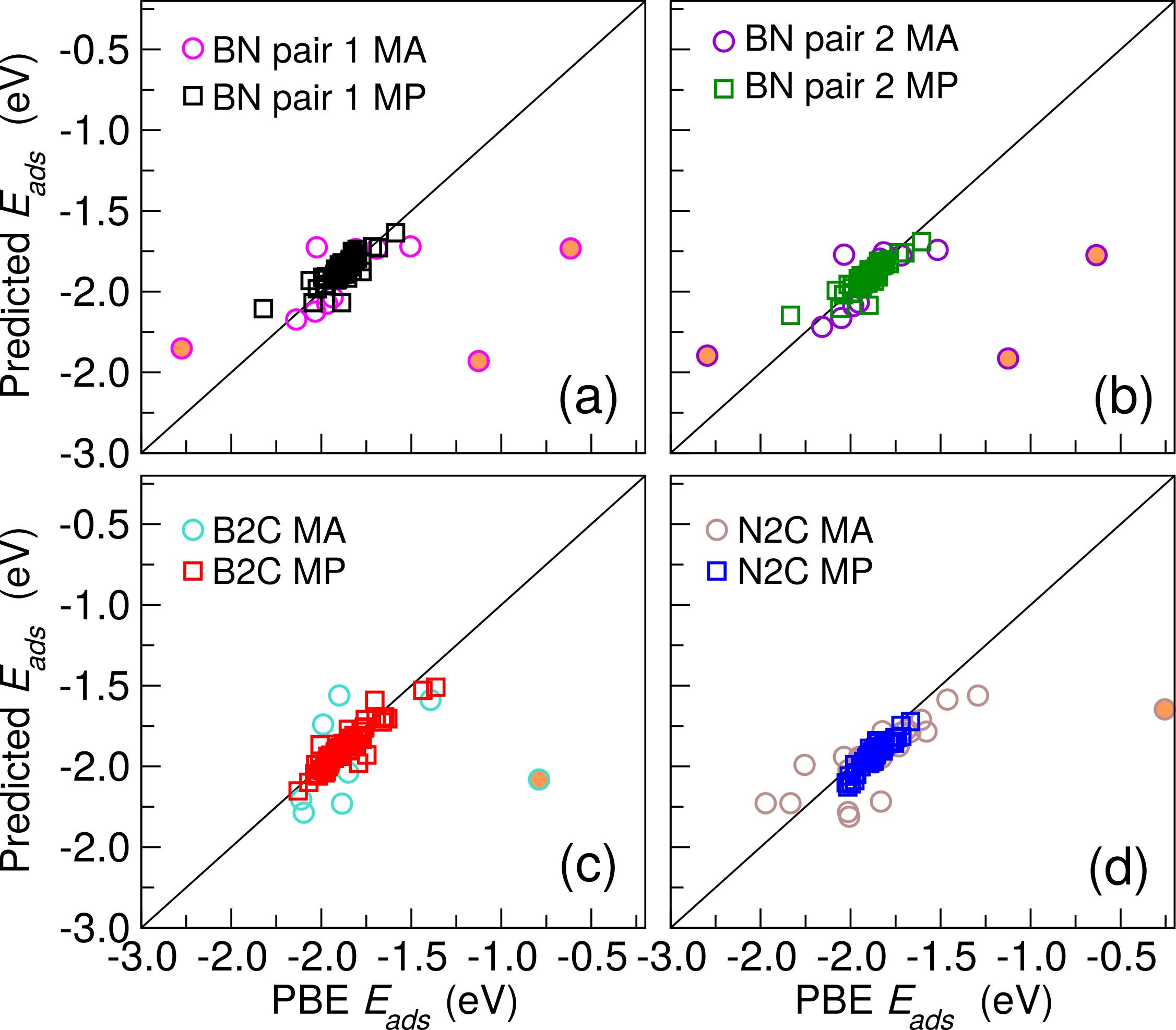} \caption{Calculated PBE
dissociative adsorption energies of water against the alchemically predicted
energies for paths in BN pair 1 (a), BN pair 2 (b), B2C (c), and N2C (d).
Squares correspond to mesomerically passive sites in the initial state, and
circles correspond to mesomerically active sites in the initial state. Filled
orange symbols indicate outliers.  }\label{figure5} \end{figure}

The use of HOMO charges can be demonstrated by comparing the quality of
predictions for two sets of paths, defined by a cut-off in their combined HOMO
charges. More specifically, paths with a combined HOMO charge higher than a
given cut-off charge are referred to as mesomerically active, and those with a
lower charge are referred to as mesomerically passive.  Here, the cut-off charge
is chosen as the lowest combined HOMO charge found in paths with a RAE $>0.1$.
In this way, we knowingly class all paths with RAE $>0.1$ as mesomerically
active, and paths with RAE less than 0.1 are classed as mesomerically passive.
Using this hindsight classification, the cut-off charges for the four routes are
0.203, 0.192, 0.140, and 0.025 e/atom for BN pair 1, BN pair 2, B2C, and N2C,
respectively. Later we discuss how the cut-off charge can be chosen \textit{a
priori} without a threshold RAE, but its usefulness is first demonstrated in
Fig.~\ref{figure5}. It can be seen that all outliers belong to the mesomerically
active paths (see filled circles in Fig.~\ref{figure5}). In addition, the
mesomerically passive paths deviate less from the PBE calculated energies and
are therefore better predicted than mesomerically active paths.

The effectiveness of this procedure is more clearly seen in Table \ref{table_1}
where the R$^2$, Spearman's rank coefficient (\rsc), mean absolute errors (MAE)
and root mean square errors (RMSE) are reported for each route. The MAE and RMSE
are an order of magnitude larger for paths involving mesomerically active sites
compared to passive sites. The MAE for mesomerically passive sites is \s0.03 eV
for the BN pair routes and thus within the so called chemical accuracy (\s0.04
eV) of the PBE adsorption energies. Similarly, the MAE for mesomerically passive
paths in the B2C and N2C routes is only slightly larger (\s0.05 eV).
Interestingly, the errors are generally larger for N2C (see in
Fig.~\ref{figure4}(b) the comparison with B2C) and as a result a smaller charge
cut-off was used based on the threshold RAE of 0.1. The larger errors for N2C
may seem at odds with the very good \rsc\ coefficient for both mesomerically
active (0.89) and passive (0.92) sites. Indeed from Fig.~\ref{figure5}(d), it
can be seen that there is only one obvious outlier in the N2C route. However, it
has been shown previously that predictions for right-to-left transformations in
the periodic table are not equivalent to the reverse and entail larger errors
\cite{chang2016fast}. We see this in the N2C route in which a nitrogen atom
takes the place of different carbon atoms across the surface. Encouragingly, a
strong correlation is still present between alchemically predicted and PBE
calculated adsorption energies in the N2C route, despite a general shift away
from the calculated energies. 
\begin{table*}[ht]
%\begin{ruledtabular}
\centering
\caption{Statistical analysis of data from four types of alchemical
  routes and resolved for mesomerically active (MA) and passive (MP)
  classification using threshold HOMO charges for each
  route. The threshold charges used for BN pair 1, BN pair 2, B2C and
  N2C are 0.203, 0.192, 0.140 and 0.025 e/atom, respectively. R$^2$
  coefficient, Spearman's rank correlation coefficient (\rsc), mean
  absolute error (MAE) in eV, and root mean square error (RMSE) in eV
  are listed. Numbers in parentheses specify the number of paths.}
\label{table_1}
\begin{tabular*}{1.0\textwidth}{@{\extracolsep{\fill} }lcccc}
\hline
          & R$^2$   & \rsc & MAE (eV)  & RMSE (eV) \\ \hline
\textit{BN pair 1} &      &   &      &      \\
total     & 0.14 &      & 0.068 & 0.193 \\
MA (38)       & 0.10 & 0.46 & 0.296 & 0.504 \\
MP (56)       & 0.72 & 0.85 & 0.032 & 0.048 \\
&      &      &       &       \\
\textit{BN pair 2} &      &   &      &      \\
total     & 0.17 &      & 0.063 & 0.191 \\
MA (42)       & 0.12 & 0.48 & 0.280 & 0.485 \\
MP (52)       & 0.79 & 0.89 & 0.025 & 0.041 \\
&      &      &       &       \\
\textit{B2C}  &      &   &      &      \\
total     & 0.27 &      & 0.068 & 0.193 \\
MA (20)   & 0.03 & 0.33 & 0.318 & 0.505 \\
MP (74)   & 0.85 & 0.88 & 0.045 & 0.130 \\
&      &      &       &       \\
\textit{N2C}  &      &   &      &      \\
total        & 0.49 &      & 0.091 & 0.176 \\
MA (20)      & 0.49 & 0.89 & 0.168 & 0.298 \\
MP (74)      & 0.87 & 0.92 & 0.055 & 0.061 \\ \hline
\end{tabular*}
\end{table*}

\section{Discussion}\label{discussion}
Partitioning the HOMO charge density for the initial states is shown to be an
effective means of filtering out particularly weak predictions. However, at
least two important questions need to be addressed with regards to this process
and let us also draw some chemical insights.

First, how should the initial threshold value for the HOMO charge density be
chosen without performing further calculations? This is somewhat of an arbitrary
choice but some guidelines can be used. For example, the threshold charge can be
chosen by considering the distribution of combined HOMO charges of all paths,
and finding the point at which the combined HOMO charge begins to deviate from
the majority of paths. For example, without considering the RAE and focusing
only on the spread of values in the combined Bader charge in Fig. \ref{figure4},
most combined charges are below 0.15 e/atom. Importantly, this choice does not rely
on a knowledge of the direct PBE results and interestingly, it is comparable to
the informed choice of threshold values for the BN pair and B2C routes in Table
\ref{table_1}. Indeed, according to Fig. \ref{figure4} a threshold value of 0.15
e/atom would still correspond to small errors for all routes considered.

Second, how can the filtered mesomerically active paths be salvaged? In the
current context that would be very useful because the most negative dissociation
energies arise from doping at mesomerically active sites (see Fig.
\ref{figure4}). Two particular solutions can be pursued. One is to simply
perform DFT calculations for the mesomerically active paths - this is somewhat
unimaginative but straightforward. The second possibility is to go beyond the
first order alchemical derivative, and improve the prediction by including
second order terms. Recently Chang \etal\ compared three approximations to the
second order term namely, the coupled perturbed (CP) approach, the independent
particle approximation (IPA) and the finite difference method, for the density
response to alchemical coupling \cite{chang2016fast}.  The CP approach is shown
to be superior to IPA for horizontal isoelectronic transformations in
many-electron systems. However, all higher order alchemical derivative terms
require additional computational cost. As such, it depends on the implementation
of second order derivative approaches whether they would be more efficient than
to directly calculate the DFT energies for mesomerically active paths.

\begin{figure} \centering \includegraphics[width=0.50\textwidth]{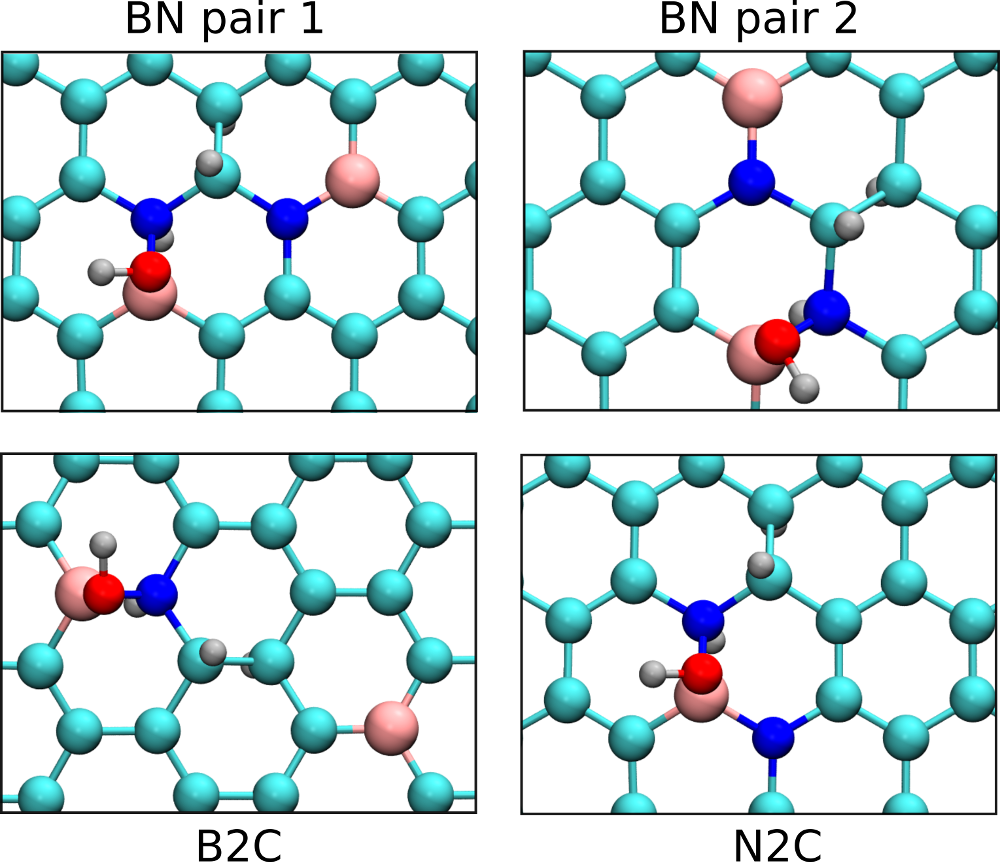}
\caption{Favorable configurations of dissociative adsorption on BN doped
graphene found in this study from alchemical screening and confirmed by direct
PBE calculations. The configuration with the most negative adsorption energy
from each alchemical route is shown. The exact PBE adsorption energies are
$-2.8$ eV for the BN pair paths, $-2.5$ eV for the N2C path, and $-2.1$ eV for
the B2C path. } \label{figure6} \end{figure}

Beyond the implications of efficiently screening isoelectronically doped
configurations of graphene, one can take a closer look at the resulting
favorable dissociative adsorption configurations to gain some chemical insight.
Fig. \ref{figure6} shows the configuration with the most favorable water
adsorption energies obtained from alchemical predictions as well as direct PBE
calculations, for each route, which range from $-2.1$ to $-2.8$ eV. Despite
different starting sublattices for BN pair 1 and BN pair 2, the same
configuration is identified as the most favorable for water dissociation. In
this state, the hydrogen atom of water adsorbs on a carbon atom between two
nitrogen atoms. This is not surprising given that the central carbon atom
becomes more positive as a result of the electronegative nitrogen atoms, and is
stabilized by bonding to a hydrogen atom. This is in agreement with the patterns
identified in a previous DFT study\cite{al2016tuning}. Similarly, in the
configuration found for N2C, the boron atom is between two nitrogen atoms and
thus forms a stronger bond with the OH fragment of water. More interestingly,
the favorable configuration from B2C is less intuitive, with a boron atom that
is not directly bonded to a surface atom at the active site (see Fig.
\ref{figure6}) and yet it corresponds to an adsorption energy of $-2.1$ eV. The
reason behind the large negative adsorption energy for this peculiar
configuration is still not fully understood and highlights the usefulness of
screening through various topological possibilities. Whilst these adsorption
energies are large and exothermic, indicating that they are likely to be
observed in experiments, it is nonetheless important to also compute activation
barriers. Note that although this is outside the scope of this study, alchemical
derivatives can be used to predict activation barriers as previously
shown\cite{sheppard2010alchemical}. Let us also note that the most unfavorable
adsorption configurations found through the alchemical screening involve
nitrogen-nitrogen single covalent bonds in the surface, suggesting that water
adsorption on such sites is particularly unfavorable.

The exothermic adsorption energies of a water molecule on BN doped graphene
found in this study span a remarkably large range ($>1$ eV) as a result of BN
doping at various sites in the surrounding graphene sheet. Given that BN of
doping graphene has been achieved experimentally\cite{synBNDG}, and in-plane
mixtures of graphene and h-BN have also been produced\cite{hybBNG1,hybBNG}, it
would be particularly interesting to verify our findings with experiments. In
addition, hydrogenation of graphene has also been experimentally achieved (see
\textit{e.g.}  Ref. \citenum{balog2010}), and thus it is timely to explore the
thermochemistry at BN doped and hydrogenated graphene surfaces with adsorption
measurements and surface studies.

\section{Conclusion}\label{alchemconc} It has been shown that predictions using
alchemical derivatives in DFT can be used to explore the impact of isoelectronic
doping in activated graphene on the dissociative adsorption of water. Doping at
different sites around the adsorption site in the substrate leads to a spread of
$\sim$2 eV in the adsorption energy. Such a wide spread of adsorption energies
shows that doping away from the dissociative adsorption site on graphene can
have a significant impact on the adsorption energy of water. This suggests that
BN doping of graphene could be a potential method for tuning surface reactions.
Importantly, it has been demonstrated that poor alchemical predictions can be
filtered out by identifying mesomerically active and passive sites using a Bader
analysis of the HOMO in the initial state. In this way, one can efficiently
screen through the majority of configurations with very good accuracy. For
instance, in this study the MAE is as low as 0.025 eV in the dissociative
adsorption energy of water. This corresponds to less than $1\%$ error for
hundreds of PBE dissociative adsorption energies using minimal computational
effort (eight self-consistent field DFT calculations). The use of alchemical
derivatives for screening can also provide an efficient way to study the
adsorption of various industrially important molecules such as hydrogen and
methane on doped graphene surfaces. More broadly, there is scope for going
beyond the first step in this study and screening materials and adsorbates in
complex catalytic processes with alchemical derivatives. Such pre-screening
could significantly reduce the number of DFT calculations that would need to be
performed whilst providing useful chemical insight.

\acknowledgments We are grateful for support from University College London and
Argonne National Laboratory (ANL) through the Thomas Young Centre-ANL
initiative. O.A.v.L. acknowledges funding from the Swiss National Science
foundation (No. PP00P2$\_$138932).  This research was partly supported by the
NCCR MARVEL, funded by the Swiss National Science Foundation.  Some of the
research leading to these results has received funding from the European
Research Council under the European Union's Seventh Framework Programme
(FP/2007-2013) / ERC Grant Agreement number 616121 (HeteroIce project). A.M. is
supported by the Royal Society through a Wolfson Research Merit Award. In
addition, we are grateful for computing resources provided by the London Centre
for Nanotechnology and Research Computing at University College London.

\bibliographystyle{apsrev4-1}\bibliography{library}

\begin{thebibliography}{69}%
\makeatletter
\providecommand \@ifxundefined [1]{%
 \@ifx{#1\undefined}
}%
\providecommand \@ifnum [1]{%
 \ifnum #1\expandafter \@firstoftwo
 \else \expandafter \@secondoftwo
 \fi
}%
\providecommand \@ifx [1]{%
 \ifx #1\expandafter \@firstoftwo
 \else \expandafter \@secondoftwo
 \fi
}%
\providecommand \natexlab [1]{#1}%
\providecommand \enquote  [1]{``#1''}%
\providecommand \bibnamefont  [1]{#1}%
\providecommand \bibfnamefont [1]{#1}%
\providecommand \citenamefont [1]{#1}%
\providecommand \href@noop [0]{\@secondoftwo}%
\providecommand \href [0]{\begingroup \@sanitize@url \@href}%
\providecommand \@href[1]{\@@startlink{#1}\@@href}%
\providecommand \@@href[1]{\endgroup#1\@@endlink}%
\providecommand \@sanitize@url [0]{\catcode `\\12\catcode `\$12\catcode
  `\&12\catcode `\#12\catcode `\^12\catcode `\_12\catcode `\%12\relax}%
\providecommand \@@startlink[1]{}%
\providecommand \@@endlink[0]{}%
\providecommand \url  [0]{\begingroup\@sanitize@url \@url }%
\providecommand \@url [1]{\endgroup\@href {#1}{\urlprefix }}%
\providecommand \urlprefix  [0]{URL }%
\providecommand \Eprint [0]{\href }%
\providecommand \doibase [0]{http://dx.doi.org/}%
\providecommand \selectlanguage [0]{\@gobble}%
\providecommand \bibinfo  [0]{\@secondoftwo}%
\providecommand \bibfield  [0]{\@secondoftwo}%
\providecommand \translation [1]{[#1]}%
\providecommand \BibitemOpen [0]{}%
\providecommand \bibitemStop [0]{}%
\providecommand \bibitemNoStop [0]{.\EOS\space}%
\providecommand \EOS [0]{\spacefactor3000\relax}%
\providecommand \BibitemShut  [1]{\csname bibitem#1\endcsname}%
\let\auto@bib@innerbib\@empty
%</preamble>
\bibitem [{\citenamefont {Curtarolo}\ \emph {et~al.}(2013)\citenamefont
  {Curtarolo}, \citenamefont {Hart}, \citenamefont {Nardelli}, \citenamefont
  {Mingo}, \citenamefont {Sanvito},\ and\ \citenamefont
  {Levy}}]{curtarolo2013}%
  \BibitemOpen
  \bibfield  {author} {\bibinfo {author} {\bibfnamefont {S.}~\bibnamefont
  {Curtarolo}}, \bibinfo {author} {\bibfnamefont {G.~L.}\ \bibnamefont {Hart}},
  \bibinfo {author} {\bibfnamefont {M.~B.}\ \bibnamefont {Nardelli}}, \bibinfo
  {author} {\bibfnamefont {N.}~\bibnamefont {Mingo}}, \bibinfo {author}
  {\bibfnamefont {S.}~\bibnamefont {Sanvito}}, \ and\ \bibinfo {author}
  {\bibfnamefont {O.}~\bibnamefont {Levy}},\ }\href@noop {} {\bibfield
  {journal} {\bibinfo  {journal} {Nat. Mater.}\ }\textbf {\bibinfo {volume}
  {12}},\ \bibinfo {pages} {191} (\bibinfo {year} {2013})}\BibitemShut
  {NoStop}%
\bibitem [{\citenamefont {Pyzer-Knapp}\ \emph {et~al.}(2015)\citenamefont
  {Pyzer-Knapp}, \citenamefont {Suh}, \citenamefont {G{\'o}mez-Bombarelli},
  \citenamefont {Aguilera-Iparraguirre},\ and\ \citenamefont
  {Aspuru-Guzik}}]{pyzer2015}%
  \BibitemOpen
  \bibfield  {author} {\bibinfo {author} {\bibfnamefont {E.~O.}\ \bibnamefont
  {Pyzer-Knapp}}, \bibinfo {author} {\bibfnamefont {C.}~\bibnamefont {Suh}},
  \bibinfo {author} {\bibfnamefont {R.}~\bibnamefont {G{\'o}mez-Bombarelli}},
  \bibinfo {author} {\bibfnamefont {J.}~\bibnamefont {Aguilera-Iparraguirre}},
  \ and\ \bibinfo {author} {\bibfnamefont {A.}~\bibnamefont {Aspuru-Guzik}},\
  }\href@noop {} {\bibfield  {journal} {\bibinfo  {journal} {Annu. Rev. Mater.
  Res.}\ }\textbf {\bibinfo {volume} {45}},\ \bibinfo {pages} {195} (\bibinfo
  {year} {2015})}\BibitemShut {NoStop}%
\bibitem [{\citenamefont {Pfeif}\ and\ \citenamefont
  {Kroenlein}(2016)}]{pfeif2016}%
  \BibitemOpen
  \bibfield  {author} {\bibinfo {author} {\bibfnamefont {E.}~\bibnamefont
  {Pfeif}}\ and\ \bibinfo {author} {\bibfnamefont {K.}~\bibnamefont
  {Kroenlein}},\ }\href@noop {} {\bibfield  {journal} {\bibinfo  {journal} {APL
  Materials}\ }\textbf {\bibinfo {volume} {4}},\ \bibinfo {pages} {053203}
  (\bibinfo {year} {2016})}\BibitemShut {NoStop}%
\bibitem [{\citenamefont {Faber}\ \emph {et~al.}(2016)\citenamefont {Faber},
  \citenamefont {Lindmaa}, \citenamefont {von Lilienfeld},\ and\ \citenamefont
  {Armiento}}]{faber2016}%
  \BibitemOpen
  \bibfield  {author} {\bibinfo {author} {\bibfnamefont {F.~A.}\ \bibnamefont
  {Faber}}, \bibinfo {author} {\bibfnamefont {A.}~\bibnamefont {Lindmaa}},
  \bibinfo {author} {\bibfnamefont {O.~A.}\ \bibnamefont {von Lilienfeld}}, \
  and\ \bibinfo {author} {\bibfnamefont {R.}~\bibnamefont {Armiento}},\
  }\href@noop {} {\bibfield  {journal} {\bibinfo  {journal} {Phys. Rev. Lett.}\
  }\textbf {\bibinfo {volume} {117}},\ \bibinfo {pages} {135502} (\bibinfo
  {year} {2016})}\BibitemShut {NoStop}%
\bibitem [{\citenamefont {Weymuth}\ and\ \citenamefont
  {Reiher}(2014)}]{weymouth2014}%
  \BibitemOpen
  \bibfield  {author} {\bibinfo {author} {\bibfnamefont {T.}~\bibnamefont
  {Weymuth}}\ and\ \bibinfo {author} {\bibfnamefont {M.}~\bibnamefont
  {Reiher}},\ }\href@noop {} {\bibfield  {journal} {\bibinfo  {journal} {Int.
  J. Quantum Chem.}\ }\textbf {\bibinfo {volume} {114}},\ \bibinfo {pages}
  {823} (\bibinfo {year} {2014})}\BibitemShut {NoStop}%
\bibitem [{\citenamefont {Cerqueira}\ \emph {et~al.}(2015)\citenamefont
  {Cerqueira}, \citenamefont {Sarmiento-P\'{e}rez}, \citenamefont {Amsler},
  \citenamefont {Nogueira}, \citenamefont {Botti},\ and\ \citenamefont
  {Marques}}]{cerqueira2015}%
  \BibitemOpen
  \bibfield  {author} {\bibinfo {author} {\bibfnamefont {T.~F.~T.}\
  \bibnamefont {Cerqueira}}, \bibinfo {author} {\bibfnamefont {R.}~\bibnamefont
  {Sarmiento-P\'{e}rez}}, \bibinfo {author} {\bibfnamefont {M.}~\bibnamefont
  {Amsler}}, \bibinfo {author} {\bibfnamefont {F.}~\bibnamefont {Nogueira}},
  \bibinfo {author} {\bibfnamefont {S.}~\bibnamefont {Botti}}, \ and\ \bibinfo
  {author} {\bibfnamefont {M.~A.~L.}\ \bibnamefont {Marques}},\ }\href@noop {}
  {\bibfield  {journal} {\bibinfo  {journal} {J. Chem. Theory Comput.}\
  }\textbf {\bibinfo {volume} {11}},\ \bibinfo {pages} {3955} (\bibinfo {year}
  {2015})}\BibitemShut {NoStop}%
\bibitem [{\citenamefont {Greeley}\ \emph {et~al.}(2006)\citenamefont
  {Greeley}, \citenamefont {Jaramillo}, \citenamefont {Bonde}, \citenamefont
  {Chorkendorff},\ and\ \citenamefont {N{\o}rskov}}]{greeley2006}%
  \BibitemOpen
  \bibfield  {author} {\bibinfo {author} {\bibfnamefont {J.}~\bibnamefont
  {Greeley}}, \bibinfo {author} {\bibfnamefont {T.~F.}\ \bibnamefont
  {Jaramillo}}, \bibinfo {author} {\bibfnamefont {J.}~\bibnamefont {Bonde}},
  \bibinfo {author} {\bibfnamefont {I.}~\bibnamefont {Chorkendorff}}, \ and\
  \bibinfo {author} {\bibfnamefont {J.~K.}\ \bibnamefont {N{\o}rskov}},\
  }\href@noop {} {\bibfield  {journal} {\bibinfo  {journal} {Nat. Mater.}\
  }\textbf {\bibinfo {volume} {5}},\ \bibinfo {pages} {909} (\bibinfo {year}
  {2006})}\BibitemShut {NoStop}%
\bibitem [{\citenamefont {Deng}\ \emph {et~al.}(2016)\citenamefont {Deng},
  \citenamefont {Novoselov}, \citenamefont {Fu}, \citenamefont {Zheng},
  \citenamefont {Tian},\ and\ \citenamefont {Bao}}]{deng2016catalysis}%
  \BibitemOpen
  \bibfield  {author} {\bibinfo {author} {\bibfnamefont {D.}~\bibnamefont
  {Deng}}, \bibinfo {author} {\bibfnamefont {K.}~\bibnamefont {Novoselov}},
  \bibinfo {author} {\bibfnamefont {Q.}~\bibnamefont {Fu}}, \bibinfo {author}
  {\bibfnamefont {N.}~\bibnamefont {Zheng}}, \bibinfo {author} {\bibfnamefont
  {Z.}~\bibnamefont {Tian}}, \ and\ \bibinfo {author} {\bibfnamefont
  {X.}~\bibnamefont {Bao}},\ }\href@noop {} {\bibfield  {journal} {\bibinfo
  {journal} {Nat. Nanotechnol.}\ }\textbf {\bibinfo {volume} {11}},\ \bibinfo
  {pages} {218} (\bibinfo {year} {2016})}\BibitemShut {NoStop}%
\bibitem [{\citenamefont {Li}\ \emph {et~al.}(2013)\citenamefont {Li},
  \citenamefont {Zhao},\ and\ \citenamefont {Yang}}]{li_2013semi}%
  \BibitemOpen
  \bibfield  {author} {\bibinfo {author} {\bibfnamefont {X.}~\bibnamefont
  {Li}}, \bibinfo {author} {\bibfnamefont {J.}~\bibnamefont {Zhao}}, \ and\
  \bibinfo {author} {\bibfnamefont {J.}~\bibnamefont {Yang}},\ }\href@noop {}
  {\bibfield  {journal} {\bibinfo  {journal} {Sci. Rep.}\ }\textbf {\bibinfo
  {volume} {3}},\ \bibinfo {pages} {1} (\bibinfo {year} {2013})}\BibitemShut
  {NoStop}%
\bibitem [{\citenamefont {Sun}\ \emph {et~al.}(2011)\citenamefont {Sun},
  \citenamefont {Wu},\ and\ \citenamefont {Shi}}]{sun_2011}%
  \BibitemOpen
  \bibfield  {author} {\bibinfo {author} {\bibfnamefont {Y.}~\bibnamefont
  {Sun}}, \bibinfo {author} {\bibfnamefont {Q.}~\bibnamefont {Wu}}, \ and\
  \bibinfo {author} {\bibfnamefont {G.}~\bibnamefont {Shi}},\ }\href@noop {}
  {\bibfield  {journal} {\bibinfo  {journal} {Ener. Environ. Sci.}\ }\textbf
  {\bibinfo {volume} {4}},\ \bibinfo {pages} {1113} (\bibinfo {year}
  {2011})}\BibitemShut {NoStop}%
\bibitem [{\citenamefont {Li}\ \emph {et~al.}(2011)\citenamefont {Li},
  \citenamefont {Guo}, \citenamefont {Yu}, \citenamefont {Ran}, \citenamefont
  {Zhang}, \citenamefont {Yan},\ and\ \citenamefont {Gong}}]{li2011highly}%
  \BibitemOpen
  \bibfield  {author} {\bibinfo {author} {\bibfnamefont {Q.}~\bibnamefont
  {Li}}, \bibinfo {author} {\bibfnamefont {B.}~\bibnamefont {Guo}}, \bibinfo
  {author} {\bibfnamefont {J.}~\bibnamefont {Yu}}, \bibinfo {author}
  {\bibfnamefont {J.}~\bibnamefont {Ran}}, \bibinfo {author} {\bibfnamefont
  {B.}~\bibnamefont {Zhang}}, \bibinfo {author} {\bibfnamefont
  {H.}~\bibnamefont {Yan}}, \ and\ \bibinfo {author} {\bibfnamefont {J.~R.}\
  \bibnamefont {Gong}},\ }\href@noop {} {\bibfield  {journal} {\bibinfo
  {journal} {J. Am. Chem. Soc.}\ }\textbf {\bibinfo {volume} {133}},\ \bibinfo
  {pages} {10878} (\bibinfo {year} {2011})}\BibitemShut {NoStop}%
\bibitem [{\citenamefont {Zheng}\ \emph
  {et~al.}(2013{\natexlab{a}})\citenamefont {Zheng}, \citenamefont {Jiao},
  \citenamefont {Ge}, \citenamefont {Jaroniec},\ and\ \citenamefont
  {Qiao}}]{zheng2013two}%
  \BibitemOpen
  \bibfield  {author} {\bibinfo {author} {\bibfnamefont {Y.}~\bibnamefont
  {Zheng}}, \bibinfo {author} {\bibfnamefont {Y.}~\bibnamefont {Jiao}},
  \bibinfo {author} {\bibfnamefont {L.}~\bibnamefont {Ge}}, \bibinfo {author}
  {\bibfnamefont {M.}~\bibnamefont {Jaroniec}}, \ and\ \bibinfo {author}
  {\bibfnamefont {S.~Z.}\ \bibnamefont {Qiao}},\ }\href@noop {} {\bibfield
  {journal} {\bibinfo  {journal} {Angew. Chem.}\ }\textbf {\bibinfo {volume}
  {125}},\ \bibinfo {pages} {3192} (\bibinfo {year}
  {2013}{\natexlab{a}})}\BibitemShut {NoStop}%
\bibitem [{\citenamefont {Kong}\ \emph
  {et~al.}(2014{\natexlab{a}})\citenamefont {Kong}, \citenamefont {Chen},\ and\
  \citenamefont {Chen}}]{kong2014doped}%
  \BibitemOpen
  \bibfield  {author} {\bibinfo {author} {\bibfnamefont {X.-K.}\ \bibnamefont
  {Kong}}, \bibinfo {author} {\bibfnamefont {C.-L.}\ \bibnamefont {Chen}}, \
  and\ \bibinfo {author} {\bibfnamefont {Q.-W.}\ \bibnamefont {Chen}},\
  }\href@noop {} {\bibfield  {journal} {\bibinfo  {journal} {Chem. Soc. Rev.}\
  }\textbf {\bibinfo {volume} {43}},\ \bibinfo {pages} {2841} (\bibinfo {year}
  {2014}{\natexlab{a}})}\BibitemShut {NoStop}%
\bibitem [{\citenamefont {Machado}\ and\ \citenamefont
  {Serp}(2012)}]{machado2012graphene}%
  \BibitemOpen
  \bibfield  {author} {\bibinfo {author} {\bibfnamefont {B.~F.}\ \bibnamefont
  {Machado}}\ and\ \bibinfo {author} {\bibfnamefont {P.}~\bibnamefont {Serp}},\
  }\href@noop {} {\bibfield  {journal} {\bibinfo  {journal} {Catal. Sci.
  Technol.}\ }\textbf {\bibinfo {volume} {2}},\ \bibinfo {pages} {54} (\bibinfo
  {year} {2012})}\BibitemShut {NoStop}%
\bibitem [{\citenamefont {Fan}\ \emph {et~al.}(2015)\citenamefont {Fan},
  \citenamefont {Zhang},\ and\ \citenamefont {Zhang}}]{fan2015multiple}%
  \BibitemOpen
  \bibfield  {author} {\bibinfo {author} {\bibfnamefont {X.}~\bibnamefont
  {Fan}}, \bibinfo {author} {\bibfnamefont {G.}~\bibnamefont {Zhang}}, \ and\
  \bibinfo {author} {\bibfnamefont {F.}~\bibnamefont {Zhang}},\ }\href@noop {}
  {\bibfield  {journal} {\bibinfo  {journal} {Chem. Soc. Rev.}\ }\textbf
  {\bibinfo {volume} {44}},\ \bibinfo {pages} {3023} (\bibinfo {year}
  {2015})}\BibitemShut {NoStop}%
\bibitem [{\citenamefont {Yao}\ \emph {et~al.}(2014)\citenamefont {Yao},
  \citenamefont {Fu}, \citenamefont {Zhang}, \citenamefont {Weng},
  \citenamefont {Li}, \citenamefont {Chen}, \citenamefont {Jin}, \citenamefont
  {Dong}, \citenamefont {Mu}, \citenamefont {Jiang} \emph
  {et~al.}}]{yao2014graphene}%
  \BibitemOpen
  \bibfield  {author} {\bibinfo {author} {\bibfnamefont {Y.}~\bibnamefont
  {Yao}}, \bibinfo {author} {\bibfnamefont {Q.}~\bibnamefont {Fu}}, \bibinfo
  {author} {\bibfnamefont {Y.}~\bibnamefont {Zhang}}, \bibinfo {author}
  {\bibfnamefont {X.}~\bibnamefont {Weng}}, \bibinfo {author} {\bibfnamefont
  {H.}~\bibnamefont {Li}}, \bibinfo {author} {\bibfnamefont {M.}~\bibnamefont
  {Chen}}, \bibinfo {author} {\bibfnamefont {L.}~\bibnamefont {Jin}}, \bibinfo
  {author} {\bibfnamefont {A.}~\bibnamefont {Dong}}, \bibinfo {author}
  {\bibfnamefont {R.}~\bibnamefont {Mu}}, \bibinfo {author} {\bibfnamefont
  {P.}~\bibnamefont {Jiang}},  \emph {et~al.},\ }\href@noop {} {\bibfield
  {journal} {\bibinfo  {journal} {Proc. Natl. Acad. Sci.}\ }\textbf {\bibinfo
  {volume} {111}},\ \bibinfo {pages} {17023} (\bibinfo {year}
  {2014})}\BibitemShut {NoStop}%
\bibitem [{\citenamefont {Grosjean}\ \emph {et~al.}(2016)\citenamefont
  {Grosjean}, \citenamefont {P{\'{e}}an}, \citenamefont {Siria}, \citenamefont
  {Bocquet}, \citenamefont {Vuilleumier},\ and\ \citenamefont
  {Bocquet}}]{Grosjean2016}%
  \BibitemOpen
  \bibfield  {author} {\bibinfo {author} {\bibfnamefont {B.}~\bibnamefont
  {Grosjean}}, \bibinfo {author} {\bibfnamefont {C.}~\bibnamefont
  {P{\'{e}}an}}, \bibinfo {author} {\bibfnamefont {A.}~\bibnamefont {Siria}},
  \bibinfo {author} {\bibfnamefont {L.}~\bibnamefont {Bocquet}}, \bibinfo
  {author} {\bibfnamefont {R.}~\bibnamefont {Vuilleumier}}, \ and\ \bibinfo
  {author} {\bibfnamefont {M.-L.}\ \bibnamefont {Bocquet}},\ }\href@noop {}
  {\bibfield  {journal} {\bibinfo  {journal} {J. Phys. Chem. Lett.}\ }\textbf
  {\bibinfo {volume} {7}},\ \bibinfo {pages} {4695} (\bibinfo {year}
  {2016})}\BibitemShut {NoStop}%
\bibitem [{\citenamefont {He}\ \emph {et~al.}(2017)\citenamefont {He},
  \citenamefont {Garnica}, \citenamefont {Bischoff}, \citenamefont {Ducke},
  \citenamefont {Bocquet}, \citenamefont {Batzill}, \citenamefont
  {Auw{\"{a}}rter},\ and\ \citenamefont {Barth}}]{He2017}%
  \BibitemOpen
  \bibfield  {author} {\bibinfo {author} {\bibfnamefont {Y.}~\bibnamefont
  {He}}, \bibinfo {author} {\bibfnamefont {M.}~\bibnamefont {Garnica}},
  \bibinfo {author} {\bibfnamefont {F.}~\bibnamefont {Bischoff}}, \bibinfo
  {author} {\bibfnamefont {J.}~\bibnamefont {Ducke}}, \bibinfo {author}
  {\bibfnamefont {M.-L.}\ \bibnamefont {Bocquet}}, \bibinfo {author}
  {\bibfnamefont {M.}~\bibnamefont {Batzill}}, \bibinfo {author} {\bibfnamefont
  {W.}~\bibnamefont {Auw{\"{a}}rter}}, \ and\ \bibinfo {author} {\bibfnamefont
  {J.~V.}\ \bibnamefont {Barth}},\ }\href {http://dx.doi.org/10.1038/nchem.2600
  http://10.0.4.14/nchem.2600
  http://www.nature.com/nchem/journal/v9/n1/abs/nchem.2600.html{\#}supplementary-information}
  {\bibfield  {journal} {\bibinfo  {journal} {Nat. Chem.}\ }\textbf {\bibinfo
  {volume} {9}},\ \bibinfo {pages} {33} (\bibinfo {year} {2017})}\BibitemShut
  {NoStop}%
\bibitem [{\citenamefont {Lattelais}\ and\ \citenamefont
  {Bocquet}(2015)}]{Lattelais2015}%
  \BibitemOpen
  \bibfield  {author} {\bibinfo {author} {\bibfnamefont {M.}~\bibnamefont
  {Lattelais}}\ and\ \bibinfo {author} {\bibfnamefont {M.~L.}\ \bibnamefont
  {Bocquet}},\ }\href@noop {} {\bibfield  {journal} {\bibinfo  {journal} {J.
  Phys. Chem. C}\ }\textbf {\bibinfo {volume} {119}},\ \bibinfo {pages} {9234}
  (\bibinfo {year} {2015})}\BibitemShut {NoStop}%
\bibitem [{\citenamefont {Altenburg}\ \emph {et~al.}(2015)\citenamefont
  {Altenburg}, \citenamefont {Lattelais}, \citenamefont {Wang}, \citenamefont
  {Bocquet},\ and\ \citenamefont {Berndt}}]{Altenburg2015}%
  \BibitemOpen
  \bibfield  {author} {\bibinfo {author} {\bibfnamefont {S.~J.}\ \bibnamefont
  {Altenburg}}, \bibinfo {author} {\bibfnamefont {M.}~\bibnamefont
  {Lattelais}}, \bibinfo {author} {\bibfnamefont {B.}~\bibnamefont {Wang}},
  \bibinfo {author} {\bibfnamefont {M.~L.}\ \bibnamefont {Bocquet}}, \ and\
  \bibinfo {author} {\bibfnamefont {R.}~\bibnamefont {Berndt}},\ }\href@noop {}
  {\bibfield  {journal} {\bibinfo  {journal} {J. Am. Chem. Soc.}\ }\textbf
  {\bibinfo {volume} {137}},\ \bibinfo {pages} {9452} (\bibinfo {year}
  {2015})}\BibitemShut {NoStop}%
\bibitem [{\citenamefont {Al-Hamdani}\ \emph {et~al.}(2016)\citenamefont
  {Al-Hamdani}, \citenamefont {Alf{\`e}}, \citenamefont {von Lilienfeld},\ and\
  \citenamefont {Michaelides}}]{al2016tuning}%
  \BibitemOpen
  \bibfield  {author} {\bibinfo {author} {\bibfnamefont {Y.~S.}\ \bibnamefont
  {Al-Hamdani}}, \bibinfo {author} {\bibfnamefont {D.}~\bibnamefont
  {Alf{\`e}}}, \bibinfo {author} {\bibfnamefont {O.~A.}\ \bibnamefont {von
  Lilienfeld}}, \ and\ \bibinfo {author} {\bibfnamefont {A.}~\bibnamefont
  {Michaelides}},\ }\href@noop {} {\bibfield  {journal} {\bibinfo  {journal}
  {J. Chem. Phys.}\ }\textbf {\bibinfo {volume} {144}},\ \bibinfo {pages}
  {154706} (\bibinfo {year} {2016})}\BibitemShut {NoStop}%
\bibitem [{\citenamefont {Sheng}\ \emph {et~al.}(2011)\citenamefont {Sheng},
  \citenamefont {Shao}, \citenamefont {Chen}, \citenamefont {Bao},
  \citenamefont {Wang},\ and\ \citenamefont {Xia}}]{Sheng2011}%
  \BibitemOpen
  \bibfield  {author} {\bibinfo {author} {\bibfnamefont {Z.~H.}\ \bibnamefont
  {Sheng}}, \bibinfo {author} {\bibfnamefont {L.}~\bibnamefont {Shao}},
  \bibinfo {author} {\bibfnamefont {J.~J.}\ \bibnamefont {Chen}}, \bibinfo
  {author} {\bibfnamefont {W.~J.}\ \bibnamefont {Bao}}, \bibinfo {author}
  {\bibfnamefont {F.~B.}\ \bibnamefont {Wang}}, \ and\ \bibinfo {author}
  {\bibfnamefont {X.~H.}\ \bibnamefont {Xia}},\ }\href@noop {} {\bibfield
  {journal} {\bibinfo  {journal} {ACS Nano}\ }\textbf {\bibinfo {volume} {5}},\
  \bibinfo {pages} {4350} (\bibinfo {year} {2011})}\BibitemShut {NoStop}%
\bibitem [{\citenamefont {Liao}\ \emph {et~al.}(2014)\citenamefont {Liao},
  \citenamefont {Peng},\ and\ \citenamefont {Liu}}]{liao2014jacs}%
  \BibitemOpen
  \bibfield  {author} {\bibinfo {author} {\bibfnamefont {L.}~\bibnamefont
  {Liao}}, \bibinfo {author} {\bibfnamefont {H.}~\bibnamefont {Peng}}, \ and\
  \bibinfo {author} {\bibfnamefont {Z.}~\bibnamefont {Liu}},\ }\href {\doibase
  10.1021/ja5048297} {\bibfield  {journal} {\bibinfo  {journal} {J. Am. Chem.
  Soc.}\ }\textbf {\bibinfo {volume} {136}},\ \bibinfo {pages} {12194}
  (\bibinfo {year} {2014})}\BibitemShut {NoStop}%
\bibitem [{\citenamefont {Kong}\ \emph
  {et~al.}(2014{\natexlab{b}})\citenamefont {Kong}, \citenamefont {Enders},
  \citenamefont {Rahman},\ and\ \citenamefont {Dowben}}]{kong2014jpcm}%
  \BibitemOpen
  \bibfield  {author} {\bibinfo {author} {\bibfnamefont {L.}~\bibnamefont
  {Kong}}, \bibinfo {author} {\bibfnamefont {A.}~\bibnamefont {Enders}},
  \bibinfo {author} {\bibfnamefont {T.~S.}\ \bibnamefont {Rahman}}, \ and\
  \bibinfo {author} {\bibfnamefont {P.~A.}\ \bibnamefont {Dowben}},\ }\href
  {http://stacks.iop.org/0953-8984/26/i=44/a=443001} {\bibfield  {journal}
  {\bibinfo  {journal} {J. Phys.: Condens. Matter}\ }\textbf {\bibinfo {volume}
  {26}},\ \bibinfo {pages} {443001} (\bibinfo {year}
  {2014}{\natexlab{b}})}\BibitemShut {NoStop}%
\bibitem [{\citenamefont {Liu}(2015)}]{liu2015nsr}%
  \BibitemOpen
  \bibfield  {author} {\bibinfo {author} {\bibfnamefont {Z.}~\bibnamefont
  {Liu}},\ }\href {\doibase 10.1093/nsr/nwv006} {\bibfield  {journal} {\bibinfo
   {journal} {Natl. Sci. Rev.}\ }\textbf {\bibinfo {volume} {2}},\ \bibinfo
  {pages} {16} (\bibinfo {year} {2015})}\BibitemShut {NoStop}%
\bibitem [{\citenamefont {Elias}\ \emph {et~al.}(2009)\citenamefont {Elias},
  \citenamefont {Nair}, \citenamefont {Mohiuddin}, \citenamefont {Morozov},
  \citenamefont {Blake}, \citenamefont {Halsall}, \citenamefont {Ferrari},
  \citenamefont {Boukhvalov}, \citenamefont {Katsnelson}, \citenamefont
  {Geim},\ and\ \citenamefont {Novoselov}}]{Elias2009}%
  \BibitemOpen
  \bibfield  {author} {\bibinfo {author} {\bibfnamefont {D.~C.}\ \bibnamefont
  {Elias}}, \bibinfo {author} {\bibfnamefont {R.~R.}\ \bibnamefont {Nair}},
  \bibinfo {author} {\bibfnamefont {T.~M.~G.}\ \bibnamefont {Mohiuddin}},
  \bibinfo {author} {\bibfnamefont {S.~V.}\ \bibnamefont {Morozov}}, \bibinfo
  {author} {\bibfnamefont {P.}~\bibnamefont {Blake}}, \bibinfo {author}
  {\bibfnamefont {M.~P.}\ \bibnamefont {Halsall}}, \bibinfo {author}
  {\bibfnamefont {A.~C.}\ \bibnamefont {Ferrari}}, \bibinfo {author}
  {\bibfnamefont {D.~W.}\ \bibnamefont {Boukhvalov}}, \bibinfo {author}
  {\bibfnamefont {M.~I.}\ \bibnamefont {Katsnelson}}, \bibinfo {author}
  {\bibfnamefont {A.~K.}\ \bibnamefont {Geim}}, \ and\ \bibinfo {author}
  {\bibfnamefont {K.~S.}\ \bibnamefont {Novoselov}},\ }\href
  {http://science.sciencemag.org/content/323/5914/610.abstract} {\bibfield
  {journal} {\bibinfo  {journal} {Science}\ }\textbf {\bibinfo {volume}
  {323}},\ \bibinfo {pages} {610 LP } (\bibinfo {year} {2009})}\BibitemShut
  {NoStop}%
\bibitem [{\citenamefont {Jaiswal}\ \emph {et~al.}(2011)\citenamefont
  {Jaiswal}, \citenamefont {{Yi Xuan Lim}}, \citenamefont {Bao}, \citenamefont
  {Toh}, \citenamefont {Loh},\ and\ \citenamefont
  {{\"{O}}zyilmaz}}]{jaiswal2011}%
  \BibitemOpen
  \bibfield  {author} {\bibinfo {author} {\bibfnamefont {M.}~\bibnamefont
  {Jaiswal}}, \bibinfo {author} {\bibfnamefont {C.~H.}\ \bibnamefont {{Yi Xuan
  Lim}}}, \bibinfo {author} {\bibfnamefont {Q.}~\bibnamefont {Bao}}, \bibinfo
  {author} {\bibfnamefont {C.~T.}\ \bibnamefont {Toh}}, \bibinfo {author}
  {\bibfnamefont {K.~P.}\ \bibnamefont {Loh}}, \ and\ \bibinfo {author}
  {\bibfnamefont {B.}~\bibnamefont {{\"{O}}zyilmaz}},\ }\href {\doibase
  10.1021/nn102034y} {\bibfield  {journal} {\bibinfo  {journal} {ACS Nano}\
  }\textbf {\bibinfo {volume} {5}},\ \bibinfo {pages} {888} (\bibinfo {year}
  {2011})}\BibitemShut {NoStop}%
\bibitem [{\citenamefont {Luo}\ \emph {et~al.}(2009)\citenamefont {Luo},
  \citenamefont {Yu}, \citenamefont {Kim}, \citenamefont {Ni}, \citenamefont
  {You}, \citenamefont {Lim}, \citenamefont {Shen}, \citenamefont {Wang},\ and\
  \citenamefont {Lin}}]{luo2009}%
  \BibitemOpen
  \bibfield  {author} {\bibinfo {author} {\bibfnamefont {Z.}~\bibnamefont
  {Luo}}, \bibinfo {author} {\bibfnamefont {T.}~\bibnamefont {Yu}}, \bibinfo
  {author} {\bibfnamefont {K.-j.}\ \bibnamefont {Kim}}, \bibinfo {author}
  {\bibfnamefont {Z.}~\bibnamefont {Ni}}, \bibinfo {author} {\bibfnamefont
  {Y.}~\bibnamefont {You}}, \bibinfo {author} {\bibfnamefont {S.}~\bibnamefont
  {Lim}}, \bibinfo {author} {\bibfnamefont {Z.}~\bibnamefont {Shen}}, \bibinfo
  {author} {\bibfnamefont {S.}~\bibnamefont {Wang}}, \ and\ \bibinfo {author}
  {\bibfnamefont {J.}~\bibnamefont {Lin}},\ }\href {\doibase 10.1021/nn900371t}
  {\bibfield  {journal} {\bibinfo  {journal} {ACS Nano}\ }\textbf {\bibinfo
  {volume} {3}},\ \bibinfo {pages} {1781} (\bibinfo {year} {2009})}\BibitemShut
  {NoStop}%
\bibitem [{\citenamefont {Wojtaszek}\ \emph {et~al.}(2011)\citenamefont
  {Wojtaszek}, \citenamefont {Tombros}, \citenamefont {Caretta}, \citenamefont
  {van Loosdrecht},\ and\ \citenamefont {van Wees}}]{wojtaszek2011}%
  \BibitemOpen
  \bibfield  {author} {\bibinfo {author} {\bibfnamefont {M.}~\bibnamefont
  {Wojtaszek}}, \bibinfo {author} {\bibfnamefont {N.}~\bibnamefont {Tombros}},
  \bibinfo {author} {\bibfnamefont {A.}~\bibnamefont {Caretta}}, \bibinfo
  {author} {\bibfnamefont {P.~H.~M.}\ \bibnamefont {van Loosdrecht}}, \ and\
  \bibinfo {author} {\bibfnamefont {B.~J.}\ \bibnamefont {van Wees}},\ }\href
  {\doibase 10.1063/1.3638696} {\bibfield  {journal} {\bibinfo  {journal} {J.
  Appl. Phys.}\ }\textbf {\bibinfo {volume} {110}},\ \bibinfo {pages} {63715}
  (\bibinfo {year} {2011})}\BibitemShut {NoStop}%
\bibitem [{\citenamefont {Gong}\ \emph {et~al.}(2014)\citenamefont {Gong},
  \citenamefont {Shi}, \citenamefont {Zhang}, \citenamefont {Zhou},
  \citenamefont {Jung}, \citenamefont {Gao}, \citenamefont {Ma}, \citenamefont
  {Yang}, \citenamefont {Yang}, \citenamefont {You}, \citenamefont {Vajtai},
  \citenamefont {Xu}, \citenamefont {MacDonald}, \citenamefont {Yakobson},
  \citenamefont {Lou}, \citenamefont {Liu},\ and\ \citenamefont
  {Ajayan}}]{Gong2014}%
  \BibitemOpen
  \bibfield  {author} {\bibinfo {author} {\bibfnamefont {Y.}~\bibnamefont
  {Gong}}, \bibinfo {author} {\bibfnamefont {G.}~\bibnamefont {Shi}}, \bibinfo
  {author} {\bibfnamefont {Z.}~\bibnamefont {Zhang}}, \bibinfo {author}
  {\bibfnamefont {W.}~\bibnamefont {Zhou}}, \bibinfo {author} {\bibfnamefont
  {J.}~\bibnamefont {Jung}}, \bibinfo {author} {\bibfnamefont {W.}~\bibnamefont
  {Gao}}, \bibinfo {author} {\bibfnamefont {L.}~\bibnamefont {Ma}}, \bibinfo
  {author} {\bibfnamefont {Y.}~\bibnamefont {Yang}}, \bibinfo {author}
  {\bibfnamefont {S.}~\bibnamefont {Yang}}, \bibinfo {author} {\bibfnamefont
  {G.}~\bibnamefont {You}}, \bibinfo {author} {\bibfnamefont {R.}~\bibnamefont
  {Vajtai}}, \bibinfo {author} {\bibfnamefont {Q.}~\bibnamefont {Xu}}, \bibinfo
  {author} {\bibfnamefont {A.~H.}\ \bibnamefont {MacDonald}}, \bibinfo {author}
  {\bibfnamefont {B.~I.}\ \bibnamefont {Yakobson}}, \bibinfo {author}
  {\bibfnamefont {J.}~\bibnamefont {Lou}}, \bibinfo {author} {\bibfnamefont
  {Z.}~\bibnamefont {Liu}}, \ and\ \bibinfo {author} {\bibfnamefont {P.~M.}\
  \bibnamefont {Ajayan}},\ }\href {http://dx.doi.org/10.1038/ncomms4193
  http://10.0.4.14/ncomms4193
  http://www.nature.com/articles/ncomms4193{\#}supplementary-information}
  {\bibfield  {journal} {\bibinfo  {journal} {Nat. Commun.}\ }\textbf {\bibinfo
  {volume} {5}},\ \bibinfo {pages} {3193} (\bibinfo {year} {2014})}\BibitemShut
  {NoStop}%
\bibitem [{\citenamefont {Ci}\ \emph {et~al.}(2010{\natexlab{a}})\citenamefont
  {Ci}, \citenamefont {Song}, \citenamefont {Jin}, \citenamefont {Jariwala},
  \citenamefont {Wu}, \citenamefont {Li}, \citenamefont {Srivastava},
  \citenamefont {Wang}, \citenamefont {Storr}, \citenamefont {Balicas} \emph
  {et~al.}}]{Ci_2010}%
  \BibitemOpen
  \bibfield  {author} {\bibinfo {author} {\bibfnamefont {L.}~\bibnamefont
  {Ci}}, \bibinfo {author} {\bibfnamefont {L.}~\bibnamefont {Song}}, \bibinfo
  {author} {\bibfnamefont {C.}~\bibnamefont {Jin}}, \bibinfo {author}
  {\bibfnamefont {D.}~\bibnamefont {Jariwala}}, \bibinfo {author}
  {\bibfnamefont {D.}~\bibnamefont {Wu}}, \bibinfo {author} {\bibfnamefont
  {Y.}~\bibnamefont {Li}}, \bibinfo {author} {\bibfnamefont {A.}~\bibnamefont
  {Srivastava}}, \bibinfo {author} {\bibfnamefont {Z.}~\bibnamefont {Wang}},
  \bibinfo {author} {\bibfnamefont {K.}~\bibnamefont {Storr}}, \bibinfo
  {author} {\bibfnamefont {L.}~\bibnamefont {Balicas}},  \emph {et~al.},\
  }\href@noop {} {\bibfield  {journal} {\bibinfo  {journal} {Nat. Mater.}\
  }\textbf {\bibinfo {volume} {9}},\ \bibinfo {pages} {430} (\bibinfo {year}
  {2010}{\natexlab{a}})}\BibitemShut {NoStop}%
\bibitem [{\citenamefont {Liu}\ \emph {et~al.}(2013{\natexlab{a}})\citenamefont
  {Liu}, \citenamefont {Ma}, \citenamefont {Shi}, \citenamefont {Zhou},
  \citenamefont {Gong}, \citenamefont {Lei}, \citenamefont {Yang},
  \citenamefont {Zhang}, \citenamefont {Yu}, \citenamefont {Hackenberg} \emph
  {et~al.}}]{Liu_2013}%
  \BibitemOpen
  \bibfield  {author} {\bibinfo {author} {\bibfnamefont {Z.}~\bibnamefont
  {Liu}}, \bibinfo {author} {\bibfnamefont {L.}~\bibnamefont {Ma}}, \bibinfo
  {author} {\bibfnamefont {G.}~\bibnamefont {Shi}}, \bibinfo {author}
  {\bibfnamefont {W.}~\bibnamefont {Zhou}}, \bibinfo {author} {\bibfnamefont
  {Y.}~\bibnamefont {Gong}}, \bibinfo {author} {\bibfnamefont {S.}~\bibnamefont
  {Lei}}, \bibinfo {author} {\bibfnamefont {X.}~\bibnamefont {Yang}}, \bibinfo
  {author} {\bibfnamefont {J.}~\bibnamefont {Zhang}}, \bibinfo {author}
  {\bibfnamefont {J.}~\bibnamefont {Yu}}, \bibinfo {author} {\bibfnamefont
  {K.~P.}\ \bibnamefont {Hackenberg}},  \emph {et~al.},\ }\href@noop {}
  {\bibfield  {journal} {\bibinfo  {journal} {Nature Nanotech.}\ }\textbf
  {\bibinfo {volume} {8}},\ \bibinfo {pages} {119} (\bibinfo {year}
  {2013}{\natexlab{a}})}\BibitemShut {NoStop}%
\bibitem [{\citenamefont {Zheng}\ \emph
  {et~al.}(2013{\natexlab{b}})\citenamefont {Zheng}, \citenamefont {Jiao},
  \citenamefont {Ge}, \citenamefont {Jaroniec},\ and\ \citenamefont
  {Qiao}}]{synBNDG}%
  \BibitemOpen
  \bibfield  {author} {\bibinfo {author} {\bibfnamefont {Y.}~\bibnamefont
  {Zheng}}, \bibinfo {author} {\bibfnamefont {Y.}~\bibnamefont {Jiao}},
  \bibinfo {author} {\bibfnamefont {L.}~\bibnamefont {Ge}}, \bibinfo {author}
  {\bibfnamefont {M.}~\bibnamefont {Jaroniec}}, \ and\ \bibinfo {author}
  {\bibfnamefont {S.~Z.}\ \bibnamefont {Qiao}},\ }\href@noop {} {\bibfield
  {journal} {\bibinfo  {journal} {Angew. Chem.}\ }\textbf {\bibinfo {volume}
  {125}},\ \bibinfo {pages} {3192} (\bibinfo {year}
  {2013}{\natexlab{b}})}\BibitemShut {NoStop}%
\bibitem [{\citenamefont {Von~Lilienfeld}\ and\ \citenamefont
  {Tuckerman}(2006)}]{von2006molecular}%
  \BibitemOpen
  \bibfield  {author} {\bibinfo {author} {\bibfnamefont {O.~A.}\ \bibnamefont
  {Von~Lilienfeld}}\ and\ \bibinfo {author} {\bibfnamefont {M.~E.}\
  \bibnamefont {Tuckerman}},\ }\href@noop {} {\bibfield  {journal} {\bibinfo
  {journal} {J. Chem. Phys.}\ }\textbf {\bibinfo {volume} {125}},\ \bibinfo
  {pages} {154104} (\bibinfo {year} {2006})}\BibitemShut {NoStop}%
\bibitem [{\citenamefont {Yang}\ and\ \citenamefont {Parr}(1985)}]{Yang1985}%
  \BibitemOpen
  \bibfield  {author} {\bibinfo {author} {\bibfnamefont {W.}~\bibnamefont
  {Yang}}\ and\ \bibinfo {author} {\bibfnamefont {R.~G.}\ \bibnamefont
  {Parr}},\ }\href {\doibase 10.1073/pnas.82.20.6723} {\bibfield  {journal}
  {\bibinfo  {journal} {Proc. Natl. Acad. Sci.}\ }\textbf {\bibinfo {volume}
  {82}},\ \bibinfo {pages} {6723} (\bibinfo {year} {1985})}\BibitemShut
  {NoStop}%
\bibitem [{\citenamefont {Parr}\ and\ \citenamefont {Yang}(1984)}]{Parr1984}%
  \BibitemOpen
  \bibfield  {author} {\bibinfo {author} {\bibfnamefont {R.~G.}\ \bibnamefont
  {Parr}}\ and\ \bibinfo {author} {\bibfnamefont {W.}~\bibnamefont {Yang}},\
  }\href {\doibase 10.1021/ja00326a036} {\bibfield  {journal} {\bibinfo
  {journal} {J. Am. Chem. Soc.}\ }\textbf {\bibinfo {volume} {106}},\ \bibinfo
  {pages} {4049} (\bibinfo {year} {1984})}\BibitemShut {NoStop}%
\bibitem [{\citenamefont {Geerlings}\ \emph {et~al.}(2003)\citenamefont
  {Geerlings}, \citenamefont {De~Proft},\ and\ \citenamefont
  {Langenaeker}}]{geerlings2003conceptual}%
  \BibitemOpen
  \bibfield  {author} {\bibinfo {author} {\bibfnamefont {P.}~\bibnamefont
  {Geerlings}}, \bibinfo {author} {\bibfnamefont {F.}~\bibnamefont {De~Proft}},
  \ and\ \bibinfo {author} {\bibfnamefont {W.}~\bibnamefont {Langenaeker}},\
  }\href@noop {} {\bibfield  {journal} {\bibinfo  {journal} {Chem. Rev.}\
  }\textbf {\bibinfo {volume} {103}},\ \bibinfo {pages} {1793} (\bibinfo {year}
  {2003})}\BibitemShut {NoStop}%
\bibitem [{\citenamefont {von Lilienfeld}(2013)}]{von2013first}%
  \BibitemOpen
  \bibfield  {author} {\bibinfo {author} {\bibfnamefont {O.~A.}\ \bibnamefont
  {von Lilienfeld}},\ }\href@noop {} {\bibfield  {journal} {\bibinfo  {journal}
  {Int. J. Quant. Chem.}\ }\textbf {\bibinfo {volume} {113}},\ \bibinfo {pages}
  {1676} (\bibinfo {year} {2013})}\BibitemShut {NoStop}%
\bibitem [{\citenamefont {Geerlings}\ \emph {et~al.}(2014)\citenamefont
  {Geerlings}, \citenamefont {Fias}, \citenamefont {Boisdenghien},\ and\
  \citenamefont {De~Proft}}]{geerlings2014conceptual}%
  \BibitemOpen
  \bibfield  {author} {\bibinfo {author} {\bibfnamefont {P.}~\bibnamefont
  {Geerlings}}, \bibinfo {author} {\bibfnamefont {S.}~\bibnamefont {Fias}},
  \bibinfo {author} {\bibfnamefont {Z.}~\bibnamefont {Boisdenghien}}, \ and\
  \bibinfo {author} {\bibfnamefont {F.}~\bibnamefont {De~Proft}},\ }\href@noop
  {} {\bibfield  {journal} {\bibinfo  {journal} {Chem. Soc. Rev.}\ }\textbf
  {\bibinfo {volume} {43}},\ \bibinfo {pages} {4989} (\bibinfo {year}
  {2014})}\BibitemShut {NoStop}%
\bibitem [{\citenamefont {von Lilienfeld}\ and\ \citenamefont
  {Tuckerman}(2007)}]{von2007jctc}%
  \BibitemOpen
  \bibfield  {author} {\bibinfo {author} {\bibfnamefont {O.~A.}\ \bibnamefont
  {von Lilienfeld}}\ and\ \bibinfo {author} {\bibfnamefont {M.~E.}\
  \bibnamefont {Tuckerman}},\ }\href@noop {} {\bibfield  {journal} {\bibinfo
  {journal} {J. Chem. Theory Comput.}\ }\textbf {\bibinfo {volume} {3}},\
  \bibinfo {pages} {1083} (\bibinfo {year} {2007})}\BibitemShut {NoStop}%
\bibitem [{\citenamefont {von Lilienfeld}(2009)}]{von2010jcp}%
  \BibitemOpen
  \bibfield  {author} {\bibinfo {author} {\bibfnamefont {O.~A.}\ \bibnamefont
  {von Lilienfeld}},\ }\href@noop {} {\bibfield  {journal} {\bibinfo  {journal}
  {J. Chem. Phys.}\ }\textbf {\bibinfo {volume} {131}},\ \bibinfo {eid}
  {164102} (\bibinfo {year} {2009})}\BibitemShut {NoStop}%
\bibitem [{\citenamefont {Sheppard}\ \emph {et~al.}(2010)\citenamefont
  {Sheppard}, \citenamefont {Henkelman},\ and\ \citenamefont {von
  Lilienfeld}}]{sheppard2010alchemical}%
  \BibitemOpen
  \bibfield  {author} {\bibinfo {author} {\bibfnamefont {D.}~\bibnamefont
  {Sheppard}}, \bibinfo {author} {\bibfnamefont {G.}~\bibnamefont {Henkelman}},
  \ and\ \bibinfo {author} {\bibfnamefont {O.~A.}\ \bibnamefont {von
  Lilienfeld}},\ }\href@noop {} {\bibfield  {journal} {\bibinfo  {journal} {J.
  Chem. Phys.}\ }\textbf {\bibinfo {volume} {133}},\ \bibinfo {pages} {084104}
  (\bibinfo {year} {2010})}\BibitemShut {NoStop}%
\bibitem [{\citenamefont {Marcon}\ \emph {et~al.}(2007)\citenamefont {Marcon},
  \citenamefont {von Lilienfeld},\ and\ \citenamefont {Andrienko}}]{homoBenz}%
  \BibitemOpen
  \bibfield  {author} {\bibinfo {author} {\bibfnamefont {V.}~\bibnamefont
  {Marcon}}, \bibinfo {author} {\bibfnamefont {O.~A.}\ \bibnamefont {von
  Lilienfeld}}, \ and\ \bibinfo {author} {\bibfnamefont {D.}~\bibnamefont
  {Andrienko}},\ }\href@noop {} {\bibfield  {journal} {\bibinfo  {journal} {J.
  Chem. Phys.}\ }\textbf {\bibinfo {volume} {127}},\ \bibinfo {eid} {064305}
  (\bibinfo {year} {2007})}\BibitemShut {NoStop}%
\bibitem [{\citenamefont {Balawender}\ \emph {et~al.}(2013)\citenamefont
  {Balawender}, \citenamefont {Welearegay}, \citenamefont {Lesiuk},
  \citenamefont {De~Proft},\ and\ \citenamefont
  {Geerlings}}]{balawender2013exploring}%
  \BibitemOpen
  \bibfield  {author} {\bibinfo {author} {\bibfnamefont {R.}~\bibnamefont
  {Balawender}}, \bibinfo {author} {\bibfnamefont {M.~A.}\ \bibnamefont
  {Welearegay}}, \bibinfo {author} {\bibfnamefont {M.}~\bibnamefont {Lesiuk}},
  \bibinfo {author} {\bibfnamefont {F.}~\bibnamefont {De~Proft}}, \ and\
  \bibinfo {author} {\bibfnamefont {P.}~\bibnamefont {Geerlings}},\ }\href@noop
  {} {\bibfield  {journal} {\bibinfo  {journal} {J. Chem. Theory Comput.}\
  }\textbf {\bibinfo {volume} {9}},\ \bibinfo {pages} {5327} (\bibinfo {year}
  {2013})}\BibitemShut {NoStop}%
\bibitem [{\citenamefont {Chang}\ \emph {et~al.}(2016)\citenamefont {Chang},
  \citenamefont {Fias}, \citenamefont {Ramakrishnan},\ and\ \citenamefont {von
  Lilienfeld}}]{chang2016fast}%
  \BibitemOpen
  \bibfield  {author} {\bibinfo {author} {\bibfnamefont {K.~S.}\ \bibnamefont
  {Chang}}, \bibinfo {author} {\bibfnamefont {S.}~\bibnamefont {Fias}},
  \bibinfo {author} {\bibfnamefont {R.}~\bibnamefont {Ramakrishnan}}, \ and\
  \bibinfo {author} {\bibfnamefont {O.~A.}\ \bibnamefont {von Lilienfeld}},\
  }\href@noop {} {\bibfield  {journal} {\bibinfo  {journal} {J. Chem. Phys.}\
  }\textbf {\bibinfo {volume} {144}},\ \bibinfo {pages} {174110} (\bibinfo
  {year} {2016})}\BibitemShut {NoStop}%
\bibitem [{\citenamefont {Solovyeva}\ and\ \citenamefont {von
  Lilienfeld}(2016)}]{solovyeva2016rapid}%
  \BibitemOpen
  \bibfield  {author} {\bibinfo {author} {\bibfnamefont {A.}~\bibnamefont
  {Solovyeva}}\ and\ \bibinfo {author} {\bibfnamefont {O.~A.}\ \bibnamefont
  {von Lilienfeld}},\ }\href@noop {} {\bibfield  {journal} {\bibinfo  {journal}
  {Phys. Chem. Chem. Phys.}\ }\textbf {\bibinfo {volume} {18}},\ \bibinfo
  {pages} {31078} (\bibinfo {year} {2016})}\BibitemShut {NoStop}%
\bibitem [{\citenamefont {to~Baben}\ \emph {et~al.}(2016)\citenamefont
  {to~Baben}, \citenamefont {Achenbach},\ and\ \citenamefont {von
  Lilienfeld}}]{moritz2016jcp}%
  \BibitemOpen
  \bibfield  {author} {\bibinfo {author} {\bibfnamefont {M.}~\bibnamefont
  {to~Baben}}, \bibinfo {author} {\bibfnamefont {J.~O.}\ \bibnamefont
  {Achenbach}}, \ and\ \bibinfo {author} {\bibfnamefont {O.~A.}\ \bibnamefont
  {von Lilienfeld}},\ }\href@noop {} {\bibfield  {journal} {\bibinfo  {journal}
  {J. Chem. Phys.}\ }\textbf {\bibinfo {volume} {144}},\ \bibinfo {eid}
  {104103} (\bibinfo {year} {2016})}\BibitemShut {NoStop}%
\bibitem [{\citenamefont {Weigend}(2014)}]{Weigend2014}%
  \BibitemOpen
  \bibfield  {author} {\bibinfo {author} {\bibfnamefont {F.}~\bibnamefont
  {Weigend}},\ }\href@noop {} {\bibfield  {journal} {\bibinfo  {journal} {J.
  Chem. Phys.}\ }\textbf {\bibinfo {volume} {141}},\ \bibinfo {eid} {134103}
  (\bibinfo {year} {2014})}\BibitemShut {NoStop}%
\bibitem [{\citenamefont {Weigend}\ \emph {et~al.}(2004)\citenamefont
  {Weigend}, \citenamefont {Schrodt},\ and\ \citenamefont
  {Ahlrichs}}]{weigend2004atom}%
  \BibitemOpen
  \bibfield  {author} {\bibinfo {author} {\bibfnamefont {F.}~\bibnamefont
  {Weigend}}, \bibinfo {author} {\bibfnamefont {C.}~\bibnamefont {Schrodt}}, \
  and\ \bibinfo {author} {\bibfnamefont {R.}~\bibnamefont {Ahlrichs}},\
  }\href@noop {} {\bibfield  {journal} {\bibinfo  {journal} {J. Chem. Phys.}\
  }\textbf {\bibinfo {volume} {121}},\ \bibinfo {pages} {10380} (\bibinfo
  {year} {2004})}\BibitemShut {NoStop}%
\bibitem [{\citenamefont {von Lilienfeld}\ \emph {et~al.}(2005)\citenamefont
  {von Lilienfeld}, \citenamefont {Lins},\ and\ \citenamefont
  {Rothlisberger}}]{von2005variational}%
  \BibitemOpen
  \bibfield  {author} {\bibinfo {author} {\bibfnamefont {O.~A.}\ \bibnamefont
  {von Lilienfeld}}, \bibinfo {author} {\bibfnamefont {R.~D.}\ \bibnamefont
  {Lins}}, \ and\ \bibinfo {author} {\bibfnamefont {U.}~\bibnamefont
  {Rothlisberger}},\ }\href
  {http://link.aps.org/doi/10.1103/PhysRevLett.95.153002} {\bibfield  {journal}
  {\bibinfo  {journal} {Phys. Rev. Lett.}\ }\textbf {\bibinfo {volume} {95}},\
  \bibinfo {pages} {153002} (\bibinfo {year} {2005})}\BibitemShut {NoStop}%
\bibitem [{\citenamefont {Kresse}\ and\ \citenamefont {Hafner}(1993)}]{vasp1}%
  \BibitemOpen
  \bibfield  {author} {\bibinfo {author} {\bibfnamefont {G.}~\bibnamefont
  {Kresse}}\ and\ \bibinfo {author} {\bibfnamefont {J.}~\bibnamefont
  {Hafner}},\ }\href@noop {} {\bibfield  {journal} {\bibinfo  {journal} {Phys.
  Rev. B}\ }\textbf {\bibinfo {volume} {47}},\ \bibinfo {pages} {558} (\bibinfo
  {year} {1993})}\BibitemShut {NoStop}%
\bibitem [{\citenamefont {Kresse}\ and\ \citenamefont {Hafner}(1994)}]{vasp2}%
  \BibitemOpen
  \bibfield  {author} {\bibinfo {author} {\bibfnamefont {G.}~\bibnamefont
  {Kresse}}\ and\ \bibinfo {author} {\bibfnamefont {J.}~\bibnamefont
  {Hafner}},\ }\href@noop {} {\bibfield  {journal} {\bibinfo  {journal} {Phys.
  Rev. B}\ }\textbf {\bibinfo {volume} {49}},\ \bibinfo {pages} {14251}
  (\bibinfo {year} {1994})}\BibitemShut {NoStop}%
\bibitem [{\citenamefont {Kresse}\ and\ \citenamefont
  {Furthm{\"u}ller}(1996{\natexlab{a}})}]{vasp3}%
  \BibitemOpen
  \bibfield  {author} {\bibinfo {author} {\bibfnamefont {G.}~\bibnamefont
  {Kresse}}\ and\ \bibinfo {author} {\bibfnamefont {J.}~\bibnamefont
  {Furthm{\"u}ller}},\ }\href@noop {} {\bibfield  {journal} {\bibinfo
  {journal} {{Comp. Mater. Sci.}}\ }\textbf {\bibinfo {volume} {{6}}},\
  \bibinfo {pages} {15} (\bibinfo {year} {{1996}}{\natexlab{a}})}\BibitemShut
  {NoStop}%
\bibitem [{\citenamefont {Kresse}\ and\ \citenamefont
  {Furthm{\"u}ller}(1996{\natexlab{b}})}]{vasp4}%
  \BibitemOpen
  \bibfield  {author} {\bibinfo {author} {\bibfnamefont {G.}~\bibnamefont
  {Kresse}}\ and\ \bibinfo {author} {\bibfnamefont {J.}~\bibnamefont
  {Furthm{\"u}ller}},\ }\href@noop {} {\bibfield  {journal} {\bibinfo
  {journal} {Phys. Rev. B}\ }\textbf {\bibinfo {volume} {54}},\ \bibinfo
  {pages} {11169} (\bibinfo {year} {1996}{\natexlab{b}})}\BibitemShut {NoStop}%
\bibitem [{\citenamefont {Pearson}(1986)}]{pearson1986pnas}%
  \BibitemOpen
  \bibfield  {author} {\bibinfo {author} {\bibfnamefont {R.~G.}\ \bibnamefont
  {Pearson}},\ }\href@noop {} {\bibfield  {journal} {\bibinfo  {journal} {Proc.
  Natl. Acad. Sci.}\ }\textbf {\bibinfo {volume} {83}},\ \bibinfo {pages}
  {8440} (\bibinfo {year} {1986})}\BibitemShut {NoStop}%
\bibitem [{\citenamefont {Bl\"ochl}(1994)}]{PAW_94}%
  \BibitemOpen
  \bibfield  {author} {\bibinfo {author} {\bibfnamefont {P.~E.}\ \bibnamefont
  {Bl\"ochl}},\ }\href@noop {} {\bibfield  {journal} {\bibinfo  {journal}
  {Phys. Rev. B}\ }\textbf {\bibinfo {volume} {50}},\ \bibinfo {pages} {17953}
  (\bibinfo {year} {1994})}\BibitemShut {NoStop}%
\bibitem [{\citenamefont {Kresse}\ and\ \citenamefont
  {Joubert}(1999)}]{PAW_99}%
  \BibitemOpen
  \bibfield  {author} {\bibinfo {author} {\bibfnamefont {G.}~\bibnamefont
  {Kresse}}\ and\ \bibinfo {author} {\bibfnamefont {D.}~\bibnamefont
  {Joubert}},\ }\href@noop {} {\bibfield  {journal} {\bibinfo  {journal} {Phys.
  Rev. B}\ }\textbf {\bibinfo {volume} {59}},\ \bibinfo {pages} {1758}
  (\bibinfo {year} {1999})}\BibitemShut {NoStop}%
\bibitem [{\citenamefont {Perdew}\ \emph {et~al.}(1996)\citenamefont {Perdew},
  \citenamefont {Burke},\ and\ \citenamefont {Ernzerhof}}]{PBE}%
  \BibitemOpen
  \bibfield  {author} {\bibinfo {author} {\bibfnamefont {J.~P.}\ \bibnamefont
  {Perdew}}, \bibinfo {author} {\bibfnamefont {K.}~\bibnamefont {Burke}}, \
  and\ \bibinfo {author} {\bibfnamefont {M.}~\bibnamefont {Ernzerhof}},\
  }\href@noop {} {\bibfield  {journal} {\bibinfo  {journal} {Phys. Rev. Lett.}\
  }\textbf {\bibinfo {volume} {77}},\ \bibinfo {pages} {3865} (\bibinfo {year}
  {1996})}\BibitemShut {NoStop}%
\bibitem [{\citenamefont {Becke}(1993)}]{b3lypA}%
  \BibitemOpen
  \bibfield  {author} {\bibinfo {author} {\bibfnamefont {A.~D.}\ \bibnamefont
  {Becke}},\ }\href@noop {} {\bibfield  {journal} {\bibinfo  {journal} {J.
  Chem. Phys.}\ }\textbf {\bibinfo {volume} {98}},\ \bibinfo {pages} {5648}
  (\bibinfo {year} {1993})}\BibitemShut {NoStop}%
\bibitem [{\citenamefont {Lee}\ \emph {et~al.}(1988)\citenamefont {Lee},
  \citenamefont {Yang},\ and\ \citenamefont {Parr}}]{b3lypB}%
  \BibitemOpen
  \bibfield  {author} {\bibinfo {author} {\bibfnamefont {C.}~\bibnamefont
  {Lee}}, \bibinfo {author} {\bibfnamefont {W.}~\bibnamefont {Yang}}, \ and\
  \bibinfo {author} {\bibfnamefont {R.~G.}\ \bibnamefont {Parr}},\ }\href@noop
  {} {\bibfield  {journal} {\bibinfo  {journal} {Phys. Rev. B}\ }\textbf
  {\bibinfo {volume} {37}},\ \bibinfo {pages} {785} (\bibinfo {year}
  {1988})}\BibitemShut {NoStop}%
\bibitem [{\citenamefont {Vosko}\ \emph {et~al.}(1980)\citenamefont {Vosko},
  \citenamefont {Wilk},\ and\ \citenamefont {Nusair}}]{b3lypC}%
  \BibitemOpen
  \bibfield  {author} {\bibinfo {author} {\bibfnamefont {S.~H.}\ \bibnamefont
  {Vosko}}, \bibinfo {author} {\bibfnamefont {L.}~\bibnamefont {Wilk}}, \ and\
  \bibinfo {author} {\bibfnamefont {M.}~\bibnamefont {Nusair}},\ }\href@noop {}
  {\bibfield  {journal} {\bibinfo  {journal} {Can. J. Phys.}\ }\textbf
  {\bibinfo {volume} {58}},\ \bibinfo {pages} {1200} (\bibinfo {year}
  {1980})}\BibitemShut {NoStop}%
\bibitem [{\citenamefont {Stephens}\ \emph {et~al.}(1994)\citenamefont
  {Stephens}, \citenamefont {Devlin}, \citenamefont {Chabalowski},\ and\
  \citenamefont {Frisch}}]{b3lypD}%
  \BibitemOpen
  \bibfield  {author} {\bibinfo {author} {\bibfnamefont {P.~J.}\ \bibnamefont
  {Stephens}}, \bibinfo {author} {\bibfnamefont {F.~J.}\ \bibnamefont
  {Devlin}}, \bibinfo {author} {\bibfnamefont {C.~F.}\ \bibnamefont
  {Chabalowski}}, \ and\ \bibinfo {author} {\bibfnamefont {M.~J.}\ \bibnamefont
  {Frisch}},\ }\href@noop {} {\bibfield  {journal} {\bibinfo  {journal} {J.
  Phys. Chem.}\ }\textbf {\bibinfo {volume} {98}},\ \bibinfo {pages} {11623}
  (\bibinfo {year} {1994})}\BibitemShut {NoStop}%
\bibitem [{\citenamefont {Dion}\ \emph {et~al.}(2004)\citenamefont {Dion},
  \citenamefont {Rydberg}, \citenamefont {Schr\"oder}, \citenamefont
  {Langreth},\ and\ \citenamefont {Lundqvist}}]{vdwDF}%
  \BibitemOpen
  \bibfield  {author} {\bibinfo {author} {\bibfnamefont {M.}~\bibnamefont
  {Dion}}, \bibinfo {author} {\bibfnamefont {H.}~\bibnamefont {Rydberg}},
  \bibinfo {author} {\bibfnamefont {E.}~\bibnamefont {Schr\"oder}}, \bibinfo
  {author} {\bibfnamefont {D.~C.}\ \bibnamefont {Langreth}}, \ and\ \bibinfo
  {author} {\bibfnamefont {B.~I.}\ \bibnamefont {Lundqvist}},\ }\href@noop {}
  {\bibfield  {journal} {\bibinfo  {journal} {Phys. Rev. Lett.}\ }\textbf
  {\bibinfo {volume} {92}},\ \bibinfo {pages} {246401} (\bibinfo {year}
  {2004})}\BibitemShut {NoStop}%
\bibitem [{\citenamefont {Becke}(1986)}]{B86}%
  \BibitemOpen
  \bibfield  {author} {\bibinfo {author} {\bibfnamefont {A.~D.}\ \bibnamefont
  {Becke}},\ }\href@noop {} {\bibfield  {journal} {\bibinfo  {journal} {J.
  Chem. Phys.}\ }\textbf {\bibinfo {volume} {84}},\ \bibinfo {pages} {4524}
  (\bibinfo {year} {1986})}\BibitemShut {NoStop}%
\bibitem [{\citenamefont {Klime{\v{s}}}\ \emph {et~al.}(2011)\citenamefont
  {Klime{\v{s}}}, \citenamefont {Bowler},\ and\ \citenamefont
  {Michaelides}}]{vdw_opt11}%
  \BibitemOpen
  \bibfield  {author} {\bibinfo {author} {\bibfnamefont {J.}~\bibnamefont
  {Klime{\v{s}}}}, \bibinfo {author} {\bibfnamefont {D.~R.}\ \bibnamefont
  {Bowler}}, \ and\ \bibinfo {author} {\bibfnamefont {A.}~\bibnamefont
  {Michaelides}},\ }\href@noop {} {\bibfield  {journal} {\bibinfo  {journal}
  {Phys. Rev. B}\ }\textbf {\bibinfo {volume} {83}},\ \bibinfo {pages} {195131}
  (\bibinfo {year} {2011})}\BibitemShut {NoStop}%
\bibitem [{\citenamefont {Bader}(1990)}]{bader1990atoms}%
  \BibitemOpen
  \bibfield  {author} {\bibinfo {author} {\bibfnamefont {R.~F.}\ \bibnamefont
  {Bader}},\ }\href@noop {} {\emph {\bibinfo {title} {Atoms in molecules}}}\
  (\bibinfo  {publisher} {Wiley Online Library},\ \bibinfo {year}
  {1990})\BibitemShut {NoStop}%
\bibitem [{\citenamefont {Liu}\ \emph {et~al.}(2013{\natexlab{b}})\citenamefont
  {Liu}, \citenamefont {Ma}, \citenamefont {Shi}, \citenamefont {Zhou},
  \citenamefont {Gong}, \citenamefont {Lei}, \citenamefont {Yang},
  \citenamefont {Zhang}, \citenamefont {Yu}, \citenamefont {Hackenberg} \emph
  {et~al.}}]{hybBNG1}%
  \BibitemOpen
  \bibfield  {author} {\bibinfo {author} {\bibfnamefont {Z.}~\bibnamefont
  {Liu}}, \bibinfo {author} {\bibfnamefont {L.}~\bibnamefont {Ma}}, \bibinfo
  {author} {\bibfnamefont {G.}~\bibnamefont {Shi}}, \bibinfo {author}
  {\bibfnamefont {W.}~\bibnamefont {Zhou}}, \bibinfo {author} {\bibfnamefont
  {Y.}~\bibnamefont {Gong}}, \bibinfo {author} {\bibfnamefont {S.}~\bibnamefont
  {Lei}}, \bibinfo {author} {\bibfnamefont {X.}~\bibnamefont {Yang}}, \bibinfo
  {author} {\bibfnamefont {J.}~\bibnamefont {Zhang}}, \bibinfo {author}
  {\bibfnamefont {J.}~\bibnamefont {Yu}}, \bibinfo {author} {\bibfnamefont
  {K.~P.}\ \bibnamefont {Hackenberg}},  \emph {et~al.},\ }\href@noop {}
  {\bibfield  {journal} {\bibinfo  {journal} {Nature Nanotech.}\ }\textbf
  {\bibinfo {volume} {8}},\ \bibinfo {pages} {119} (\bibinfo {year}
  {2013}{\natexlab{b}})}\BibitemShut {NoStop}%
\bibitem [{\citenamefont {Ci}\ \emph {et~al.}(2010{\natexlab{b}})\citenamefont
  {Ci}, \citenamefont {Song}, \citenamefont {Jin}, \citenamefont {Jariwala},
  \citenamefont {Wu}, \citenamefont {Li}, \citenamefont {Srivastava},
  \citenamefont {Wang}, \citenamefont {Storr}, \citenamefont {Balicas} \emph
  {et~al.}}]{hybBNG}%
  \BibitemOpen
  \bibfield  {author} {\bibinfo {author} {\bibfnamefont {L.}~\bibnamefont
  {Ci}}, \bibinfo {author} {\bibfnamefont {L.}~\bibnamefont {Song}}, \bibinfo
  {author} {\bibfnamefont {C.}~\bibnamefont {Jin}}, \bibinfo {author}
  {\bibfnamefont {D.}~\bibnamefont {Jariwala}}, \bibinfo {author}
  {\bibfnamefont {D.}~\bibnamefont {Wu}}, \bibinfo {author} {\bibfnamefont
  {Y.~e.}\ \bibnamefont {Li}}, \bibinfo {author} {\bibfnamefont
  {A.}~\bibnamefont {Srivastava}}, \bibinfo {author} {\bibfnamefont
  {Z.}~\bibnamefont {Wang}}, \bibinfo {author} {\bibfnamefont {K.}~\bibnamefont
  {Storr}}, \bibinfo {author} {\bibfnamefont {L.}~\bibnamefont {Balicas}},
  \emph {et~al.},\ }\href@noop {} {\bibfield  {journal} {\bibinfo  {journal}
  {Nat. Mater.}\ }\textbf {\bibinfo {volume} {9}},\ \bibinfo {pages} {430}
  (\bibinfo {year} {2010}{\natexlab{b}})}\BibitemShut {NoStop}%
\bibitem [{\citenamefont {Balog}\ \emph {et~al.}(2010)\citenamefont {Balog},
  \citenamefont {J{\o}rgensen}, \citenamefont {Nilsson}, \citenamefont
  {Andersen}, \citenamefont {Rienks}, \citenamefont {Bianchi}, \citenamefont
  {Fanetti}, \citenamefont {L{\ae}gsgaard}, \citenamefont {Baraldi},
  \citenamefont {Lizzit} \emph {et~al.}}]{balog2010}%
  \BibitemOpen
  \bibfield  {author} {\bibinfo {author} {\bibfnamefont {R.}~\bibnamefont
  {Balog}}, \bibinfo {author} {\bibfnamefont {B.}~\bibnamefont {J{\o}rgensen}},
  \bibinfo {author} {\bibfnamefont {L.}~\bibnamefont {Nilsson}}, \bibinfo
  {author} {\bibfnamefont {M.}~\bibnamefont {Andersen}}, \bibinfo {author}
  {\bibfnamefont {E.}~\bibnamefont {Rienks}}, \bibinfo {author} {\bibfnamefont
  {M.}~\bibnamefont {Bianchi}}, \bibinfo {author} {\bibfnamefont
  {M.}~\bibnamefont {Fanetti}}, \bibinfo {author} {\bibfnamefont
  {E.}~\bibnamefont {L{\ae}gsgaard}}, \bibinfo {author} {\bibfnamefont
  {A.}~\bibnamefont {Baraldi}}, \bibinfo {author} {\bibfnamefont
  {S.}~\bibnamefont {Lizzit}},  \emph {et~al.},\ }\href@noop {} {\bibfield
  {journal} {\bibinfo  {journal} {Nat. Mater.}\ }\textbf {\bibinfo {volume}
  {9}},\ \bibinfo {pages} {315} (\bibinfo {year} {2010})}\BibitemShut {NoStop}%
\end{thebibliography}%

\end{document}